\documentclass[aps,twocolumn,superscriptaddress,pre,superscriptreference,reprint]{revtex4-2}
\usepackage{amsmath,bbold,bm,amssymb,scalerel,mathtools}
\usepackage{graphicx}
\usepackage{color}
\usepackage{enumitem}
\usepackage{algorithm,algpseudocode}
\usepackage{multirow}
\usepackage{colortbl,booktabs}
\usepackage{placeins}
\usepackage{outlines}
\definecolor{darkblue}{rgb}{0,0,0.6}
\usepackage[colorlinks,linkcolor=darkblue,citecolor=darkblue,urlcolor=darkblue]{hyperref}

\makeatletter
\newcommand\footnoteref[1]{\protected@xdef\@thefnmark{\ref{#1}}\@footnotemark}
\makeatother

\begin{document}

\title{Roadmap on machine learning glassy dynamics}

\author{Gerhard Jung}
\affiliation{Laboratoire Charles Coulomb (L2C), Universit\'e de Montpellier, CNRS, 34095 Montpellier, France}

\affiliation{Laboratoire Interdisciplinaire de Physique (LIPhy), Université Grenoble Alpes, 38402 Saint-Martin-d'Hères, France}

\author{Rinske M. Alkemade}
\affiliation{Soft Condensed Matter and Biophysics, Debye Institute for Nanomaterials Science, Utrecht University, Utrecht, Netherlands}

\author{Victor Bapst}
\affiliation{GoogleDeepMind, London, UK}

\author{Daniele Coslovich}
\affiliation{Dipartimento di Fisica, Universit\`a di Trieste, Strada Costiera 11, 34151, Trieste, Italy}

\author{Laura Filion}
\affiliation{Soft Condensed Matter and Biophysics, Debye Institute for Nanomaterials Science, Utrecht University, Utrecht, Netherlands}

\author{Fran{\c c}ois P. Landes}
\affiliation{Universit\'e Paris-Saclay, CNRS, INRIA, Laboratoire Interdisciplinaire des Sciences du Num\'erique, TAU team, 91190 Gif-sur-Yvette, France}

\author{Andrea Liu}
\affiliation{Department of Physics and Astronomy, University of Pennsylvania, Philadelphia, PA 19104, USA}
\affiliation{Santa Fe Institute, 1399 Hyde Park Road, Santa Fe, NM 87501, USA}

\author{Francesco Saverio Pezzicoli}
\affiliation{Universit\'e Paris-Saclay, CNRS, INRIA, Laboratoire Interdisciplinaire des Sciences du Num\'erique, TAU team, 91190 Gif-sur-Yvette, France}

\author{Hayato Shiba}
\affiliation{Graduate School of Information Science, University of Hyogo, Kobe 650-0047, Japan} 

\author{Giovanni Volpe}
\affiliation{Department of Physics, University of Gothenburg, Origovägen 6B, Gothenburg 41296, Sweden}

\author{Francesco Zamponi}
\affiliation{Laboratoire de Physique de l’Ecole Normale Supérieure, ENS, Université PSL, CNRS, Sorbonne Université, Université de Paris, F-75005 Paris, France}

\author{Ludovic Berthier}
\affiliation{Laboratoire Charles Coulomb (L2C), Universit\'e de Montpellier, CNRS, 34095 Montpellier, France}

\affiliation{Gulliver, UMR CNRS 7083, ESPCI Paris, PSL Research University, 75005 Paris, France}

\author{Giulio Biroli}
\affiliation{Laboratoire de Physique de l’Ecole Normale Supérieure, ENS, Université PSL, CNRS, Sorbonne Université, Université de Paris, F-75005 Paris, France}

\date{\today}

\begin{abstract}
 Unraveling the connections between microscopic structure, emergent physical properties, and slow dynamics has long been a challenge when studying the glass transition. The absence of clear visible structural order in amorphous configurations complicates the identification of the key physical mechanisms underpinning slow dynamics. The difficulty in sampling equilibrated configurations at low temperatures hampers thorough numerical and theoretical investigations. This perspective article explores the potential of machine learning (ML) techniques to face these challenges, building on the algorithms that have revolutionized computer vision and image recognition. We present recent successful ML applications, as well as many open problems for the future, such as transferability and interpretability of ML approaches. We highlight new ideas and directions in which ML could provide breakthroughs to better understand the fundamental mechanisms at play in glass-forming liquids. To foster a collaborative community effort, this article also introduces the ``GlassBench" dataset, providing simulation data and benchmarks for both two-dimensional and three-dimensional glass-formers. We propose critical metrics to compare the performance of emerging ML methodologies, in line with benchmarking practices in image and text recognition. The goal of this roadmap is to provide guidelines for the development of ML techniques in systems displaying slow dynamics, while inspiring new directions to improve our theoretical understanding of glassy liquids. 
\end{abstract}

\maketitle

\section{Introduction}

\label{sec:intro}

 When supercooled liquids undergo a glass transition, a dramatic slowdown of transport properties is observed and the resulting material dynamically resembles a crystalline solid {---} yet one of the main characteristic of glasses is that they maintain their amorphous liquid structure~\cite{ediger1996glasses}. While glasses and other amorphous materials have been used since prehistoric times, their technological applications have blossomed in recent years~\cite{berthier2016facets}, including, e.g., metallic glasses for biomedical implants~\cite{li2016recent} and vapor-deposited organic films used extensively in visual displays~\cite{ediger2017}. Despite several decades of research involving experiments, theory and computer simulations, many fundamental mechanisms remain to be elucidated, such as macroscopic mechanical properties, highly cooperative stress relaxation in glasses, and the statistical mechanics nature of the glass transition itself~\cite{cavagna2009glasses}. 

 The raise of deep learning in the last decade~\cite{lecun2015deep} was initially driven by applications in computer vision, in particular image recognition and feature detection, which soon outperformed traditional techniques~\cite{alzubaidi2021review}. These original breakthroughs are now starting to revolutionize several other areas in technology and science. Our aim here is to address the potential of ML methods to boost research on fundamental aspects of glassy dynamics, in particular the ones that are playing an important role in advancing theories of the glass transition.

 One of the main challenges in developing a fundamental microscopic theory of glasses is the absence of any simple and visible structural order. While crystalline defects in the otherwise well-ordered structures are easily detectable, it is an open problem to find analogous structural features in amorphous materials. Over the years, many different proposals for ``defects" or locally preferred structures have been proposed and developed~\cite{Royall2015,tanaka2019revealing,aguilar2020tetrahedral,richard2020review}. This seems to indicate that, even in amorphous configurations, it could be possible to detect the emergence of some kind of short- and medium-range order. However, these identifications usually only apply to specific systems, and they are often only weakly correlated with local dynamical relaxations. There is therefore a clear need for new and more powerful system-independent ways to systematically find preferred structures in amorphous configurations. This is a challenge for which new ML methods could be a great asset, in particular thanks to the progress in unsupervised learning. Several approaches have been developed recently towards this goal~\cite{unsupervisedFilion,unsupervisedCoslovich,oyama2023deep,siavash2022markov}.

Another long-standing challenge has been understanding and characterizing the fundamental mechanisms underpinning slow and glassy dynamics, which are responsible for the glass transition. To this aim, there has been a substantial effort to identify the microscopic properties that lead to dynamical relaxation. Given a snapshot (an equilibrium configuration), several local properties have been proposed to pinpoint the regions more likely to relax within a window of, say, some fraction of the relaxation time. Examples include the local Debye-Waller factor, eigenvectors of the Hessian of inherent structures, etc. \cite{widmer2006predicting,widmer2008irreversible}. There is no consensus on what are the best predictors of future dynamics. Moreover, they could change with temperature or be system-specific, according to several theories of the glass transition~\cite{berthier_theoretical_2011}. Thanks to the advances in numerical simulations of glass-forming liquids we can now produce large datasets of initial configurations and subsequent dynamical trajectories. This provides a natural playground to apply supervised learning techniques in order to identify the local predictors of dynamical relaxation. Several researchers have taken up this challenge and developed ML methods to predict where local relaxations are more likely to take place given an initial snapshot~\cite{schoenholz2016structural, CubukPRL2015, bapst2020unveiling, Zaccone2021, Filion2021, Filion2022, GNNrelative2022,alkemade2023improving,Ciarella_2023,pezzicoli2022se,ruiz2022discovering,jung2022predicting}.

Finally, the ultimate goal of the research efforts devoted to the theory of glassy dynamics would be to 
combine the solutions of the previous problems to develop an effective theory of the glass transition. Until now, this challenge has been tackled starting from some theoretical assumptions driven by experimental and simulation results~\cite{berthier_theoretical_2011}. Also in this challenge ML methods can make a difference; they can assist in this quest by providing a complementary identification of the mechanisms inducing relaxation, as first shown in~\cite{ZhangPRR2022}.

\begin{figure}
    \centering
    \includegraphics[scale=0.7]{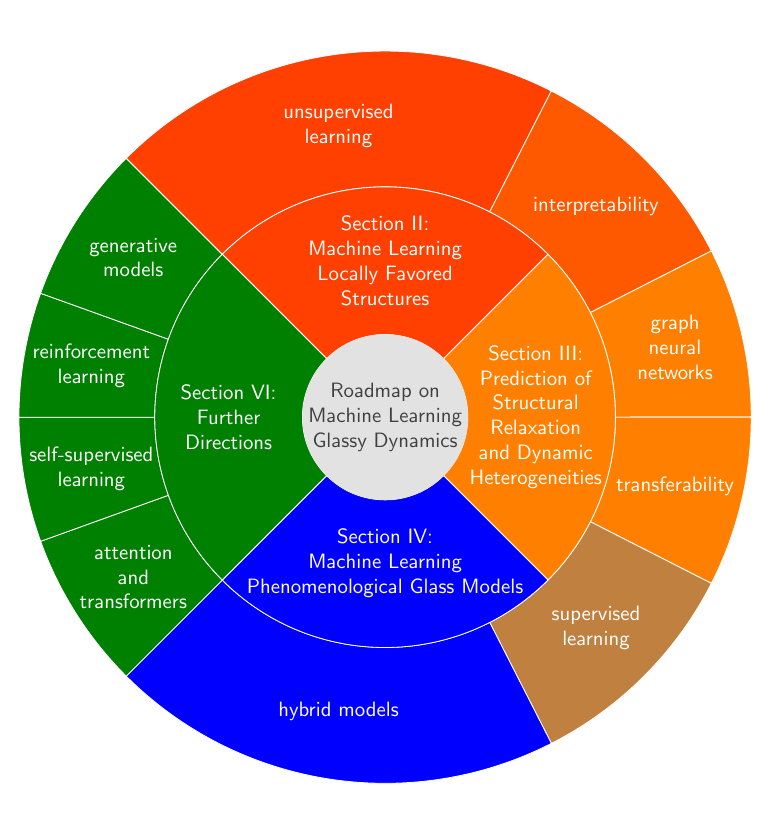}
    \caption{Visual summary of the roadmap. The individual sections at the center are connected to the big questions in the field of glass physics. They are surrounded by the various machine learning concepts used to answer them.}
    \label{fig:illustration_roadmap}
\end{figure}

Clearly, the time is ripe for investigating the ability of the new ML methods to advance our fundamental understanding of glass-forming liquids. In this roadmap, focusing on the three main goals described above, we present respectively in Secs.~\ref{sec:structure}, \ref{sec:dynamics}, \ref{sec:models} the recent contributions in this endeavor, discussing the main difficulties ahead and  the possible paths to circumvent them. In Fig.~\ref{fig:illustration_roadmap}, we give a visual overview over the different ML concepts that will be discussed within the individual sections.
In Sec.~\ref{sec:benchmark} we provide a framework ``\href{https://doi.org/10.5281/zenodo.10118191}{GlassBench}'' to enable, encourage and structure a broader community effort to further develop such ML approaches. It consists in (i)
a dataset including simulation data for a two-dimensional \cite{jung2022predicting} and a three-dimensional glass-former \cite{kob1995KA,GNNrelative2022}, (ii) benchmarks on different tasks associated to predicting local dynamics from a given initial configuration, (iii) an assessment of the current state of the art. Our purpose is to fuel and organize new developments of advanced ML techniques, as done in the field of image and text recognition, as well as generative modelling \cite{paperswithcode}.  Finally, we close this roadmap by discussing in Sec. \ref{sec:summary_directions} exciting new concepts and directions in ML which have the potential to play a very important role in future research on the theory of glassy dynamics.

\section{Machine learning locally favored structures}

\label{sec:structure}

Although glass-forming liquids and glasses lack long-range order, close inspection of their atomic structure reveals particle arrangements that are more regular, symmetric, and of lower (free-)energy than the average. Icosahedral local structures are the best known example of such favorable arrangements and they are found in several metallic alloys, colloidal suspensions as well as in computer models of glassy liquids~\cite{Royall2015}.
Such locally favored structures (LFS), distinct from the bulk of the particle arrangements and yet incompatible with crystalline order, are also key ingredients of some theoretical approaches to glass formation~\cite{berthier_theoretical_2011,tarjusViscousSlowingSupercooled2000b}. 

Despite the importance of structural analysis in glassy materials~\cite{Royall2015, tanaka2019revealing}, there is at present no generally accepted operational definition of LFS. Standard approaches, such as Voronoi tessellation~\cite{tanemuraGeometricalAnalysisCrystallization1977}, topological cluster classification~\cite{malinsIdentificationStructureCondensed2013} and other related methods~\cite{honeycuttMolecularDynamicsStudy1987,lazarTopologicalFrameworkLocal2015}, provide a detailed classification of the possible local geometric arrangements. These methods may indicate what are the most frequent or most stable local arrangements, but they are sensitive to thermal fluctuations and tend to provide a too fine-grained classification, which is difficult to exploit in a theoretical setting.
Bond-order parameters (BOP)~\cite{steinhardtBondorientationalOrderLiquids1983} provide yet another route to characterize the local structure of dense particle systems~\cite{tanaka2019revealing}. While this approach offers in principle a systematic description of the local arrangements, the choice of the relevant BOP has been traditionally guided by physical intuition~\cite{tanaka2019revealing}, which requires specific and system-dependent a priori knowledge about the relevant symmetries of the local arrangements.

Unsupervised learning methods offer natural system-independent ways to tackle
the above issues~\cite{mehtaHighbiasLowvarianceIntroduction2019,chengMappingMaterialsMolecules2020,glielmoUnsupervisedLearningMethods2021}.
Along with automated identification of phase transitions~\cite{huDiscoveringPhasesPhase2017c, rodriguez-nievaIdentifyingTopologicalOrder2019, mendes-santosUnsupervisedLearningUniversal2021b}, characterization of the properties of complex materials from high-dimensional datasets is indeed one of the key applications of unsupervised learning in condensed matter physics~\cite{chengMappingMaterialsMolecules2020}. The general idea is to first characterize a faithful, high-dimensional representation of the particle local environment based for instance on a systematic bond-order expansion of the local density~\cite{bartokRepresentingChemicalEnvironments2013} (see Ref.~\cite{parsaeifardAssessmentStructuralResolution2021a} for a recent review on structural descriptors).
Unsupervised ML methods are then used to identify a small number of collective coordinates, $\tilde{X}_i$, that account for the relevant fluctuations of the local structure, thereby reducing the dimensionality of the descriptors. Dimensionality reduction techniques range from simple principal component analysis (PCA), or its kernel variant, to more sophisticated statistical learning methods, like neural network auto-encoders (AE)~\cite{mehtaHighbiasLowvarianceIntroduction2019}.
These methods may be combined in the future with more advanced approaches, such as self-supervised learning or pre-training (see Sec.~\ref{sec:supervised_RL}), possibly exploiting the intrinsic symmetries of the system~\cite{midtvedt2022single}. 
Clustering methods can be finally applied to this reduced structural representation of the material structure, to pinpoint its heterogeneity~\cite{chengMappingMaterialsMolecules2020}.

The studies highlighted in Ref.~\cite{chengMappingMaterialsMolecules2020} focus mostly on ordered materials or disordered systems with covalent or hydrogen bonding, such as amorphous carbon~\cite{caroGrowthMechanismOrigin2018} or liquid water~\cite{monserratLiquidWaterContains2020}, where the preferred geometrical order is easy to identify thanks to low coordination numbers.
Dense amorphous systems are characterized instead by close-packed arrangements, which provide a challenging benchmark for this kind of structural analysis.
In a series of recent papers~\cite{unsupervisedFilion,unsupervisedCoslovich,coslovichDimensionalityReductionLocal2022, banerjee2023ML, banerjee2023ML2}, dimensionality reduction and clustering have been applied to models of closed-packed glass-forming liquids.
In particular, Boattini \textit{et al.}~\cite{unsupervisedFilion} have used BOP, Gaussian mixture models and a neural network AE to study glassy binary mixtures. Their work revealed a significant structural heterogeneity, suggesting that 
in these systems, we can distinguish fluctuating regions displaying two different types of local disorder. 
These regions display a connection with the temporal fluctuations of dynamics, which will be further discussed in Sec.~\ref{sec:dynamics}.

A related study by Paret {\it et al.}~\cite{unsupervisedCoslovich} addressed the issue of clustering of local structural arrangements using a different, information-theoretic approach.
At a qualitative level, the results of Refs.~\cite{unsupervisedFilion} and \cite{unsupervisedCoslovich} appear consistent with one another. However, a more recent investigation~\cite{coslovichDimensionalityReductionLocal2022} revealed a significant system-dependence of structural heterogeneity in glassy liquids. The gist of these findings is illustrated in Fig.~\ref{fig:pca}, which shows representative PCA maps obtained from a smooth bond-order (SBO) descriptor (see supplemental material (SM) for technical details~\cite{SM}). The distribution of the first two principal components is bimodal for an embedded-atom model of Cu$_{64}$Zr$_{36}$, which has a well-defined icosahedral LFS, while it is less heterogeneous for the canonical Kob-Andersen (KA) mixture (see also Sec.~\ref{sec:benchmark}), whose local arrangements display clearly different geometrical states. While these differences question the universality of the concept of LFS, the first few principal component projections always correlate with physically-motivated structural measures~\cite{coslovichDimensionalityReductionLocal2022}.
We expect that more information could be harvested by looking at chemically-resolved descriptors~\cite{offei-dansoHighDimensionalFluctuationsLiquid2022}
and on larger length scales (medium range order).
Moreover, computing the intrinsic dimension of structural datasets \cite{campadelliIntrinsicDimensionEstimation2015, mendes-santosUnsupervisedLearningUniversal2021b} may provide additional insight into the nature of structural order and its system dependence.

\begin{figure}
    \centering
    \includegraphics[]{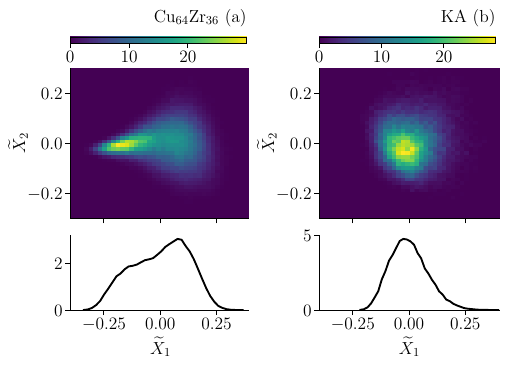}
    \caption{Principal component analysis maps of the first two principal component projections $\tilde{X}_1$ and $\tilde{X}_2$ of the smooth bond-order parameter around (a) Cu atoms in glassy Cu$_{64}$Zr$_{36}$ and (b) around small particles in the Kob-Andersen binary mixture around its mode-coupling temperature~\cite{coslovichDimensionalityReductionLocal2022}. Note the contrast between the bimodal structure in (a) due to icosahedral local structures~\cite{coslovichDimensionalityReductionLocal2022} and the homogeneous distribution of the projections in (b). The bottom panels show the marginal distribution of the first principal component projection.}
    \label{fig:pca}
\end{figure}

One of the striking observations of Ref.~\cite{coslovichDimensionalityReductionLocal2022} is that neural network AE and PCA yield identical reductions of the BOP descriptors. This suggests that either the local structure in dense glassy mixtures is simple: it displays a broad continuous spectrum of geometrical arrangements, possibly decorated by features like LFS, crystallites or structural defects, or that the current identification of LFS is missing some crucial ingredient (e.g. information about dynamics). The outcome of PCA is also easy to interpret, as the principal component directions provide direct insight into the dominant structural parameters~\footnote{Note that in the presence of agnostic, high-dimensional structural descriptors, such as the smooth atomic overlap parameters~\cite{bartokRepresentingChemicalEnvironments2013}, interpretation always occurs \textit{a posteriori}, by searching for correlations between some of the reduced structural and physically-motivated structural measures~\cite{chengMappingMaterialsMolecules2020}}.
On the one hand, these results question the utility of complex deep learning methods in studying glass structure.
On the other hand, the descriptors used in Refs.~\cite{unsupervisedFilion,unsupervisedCoslovich,coslovichDimensionalityReductionLocal2022} do not exhaust all forms of structural heterogeneity and some may be affected by some deeper shortcomings~\cite{parsaeifardManifoldsQuasiconstantSOAP2022}.
Development of structural descriptors is still very active~\cite{darbyCompressingLocalAtomic2022,darbyTensorreducedAtomicDensity2022} and these advances wait for applications in the context of glassy materials. Addressing the above issues may become crucial in future studies of more demanding benchmarks for structural characterization, such as compositional order in polydisperse glassy models~\cite{Coslovich_Ozawa_Berthier_2018, Tong_Tanaka_2023}, medium-range order in oxides or metallic glasses \cite{Elliott_1991, sheng2006atomic}, and orientational order in glassy water~\cite{Montes_de_Oca_Sciortino_Appignanesi_2020, Faccio_Benzi_Zanetti-Polzi_Daidone_2022}. 
Computational studies of these complex systems represent exciting opportunities to gain insight into the nature and role of local structure in glassy materials and to provide solid grounds for predictive theoretical approaches based on structure.

Another research line where structure-based ML approaches are making progress aims at predicting macroscopic properties of glasses relevant for applications, such as oxides or chalcogenide glasses, over a wide range of chemical compositions~\cite{mauro2016accelerating, Cassar_Mastelini_Botari_Alcobaca_deCarvalho_Zanotto_2021}.
Recent work on sodium-silicate glasses shows that physics-informed machine learning models can reliably interpolate and extrapolate these properties on the basis of structural information only~\cite{bodker2022predicting}.
These findings indicate that, despite the apparent complexity of the feature space, the relationship between local structure and macroscopic glass properties is often linear, which makes it easy for machine learning models to generalize outside their training set. See Ref.~\cite{ravinder2021artificial} for a recent roadmap covering this topic.

Having characterized amorphous structure, a crucial question is whether the structural descriptors are connected to emergent relaxation dynamics in the glass-forming liquid~\cite{doliwa2003potential,Harrowell2006,hocky2014correlation,tong2018revealing}.  As will be clear in Sec.~\ref{sec:benchmark}, current unsupervised methods provide only limited insights into the heterogeneity of the dynamics, except in specific systems dominated by strong icoshaedral order~\cite{unsupervisedFilion,unsupervisedCoslovich,coslovichDimensionalityReductionLocal2022}. 
Whether this is a technical limitation of the unsupervised methods used to date or rather an intrinsic feature of supercooled dynamics remains to be clarified.

\section{Prediction of structural relaxation and dynamic heterogeneities }

\label{sec:dynamics}

\begin{figure}
    \centering
    \includegraphics[width=\linewidth]{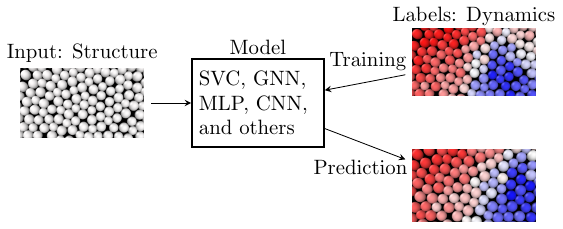}
    \caption{Typical supervised ML procedure in condensed matter. The raw input is encoded using structural descriptors or graphs. A model is trained using labels which describe structural relaxation obtained, e.g., from molecular dynamics simulations. Colors indicate frozen (blue) or rearranging (red) particles. ML techniques range from support vector classification (SVC) to graph neural networks (GNNs), multilayer perceptrons (MLPs), and convolutional neural networks (CNNs). After training, structural relaxation is predicted for previously unseen structures.}
    \label{fig:sketch}
\end{figure}

One of the central challenges for both computational and theoretical studies of glass-forming liquids is to predict future dynamics of a configuration from an initial snapshot. Note that one is not interested in predicting the whole future evolution, but only the dynamical processes leading to microscopic irreversible motion.
Supervised ML provides a natural tool to perform such prediction, essentially by fitting high-dimensional structural input to the relaxation dynamics, 
similar in spirit to classification in image recognition.   
In general, three choices need to be made to design a model:  i) which structural descriptors to use to characterize the input configuration; ii) which labels to use to quantify structural microscopic relaxation; and iii) what model and ML algorithm to use to fit the input to the labels. 
A large variety of techniques have already been introduced to tackle this problem, ranging from ridge regression using complex and coarse-grained structural descriptors to graph neural networks using raw particles positions, as illustrated in Fig.~\ref{fig:sketch}. 

In 2015, Cubuk \textit{et al.} \cite{CubukPRL2015} demonstrated that support vector classifiers (SVC)~\footnote{Note that in Ref.~\cite{CubukPRL2015} the term Support Vector Machine was used, but here we use the term Support Vector Classifier to stress the fact that the procedure corresponded to a classification and not a regression.} could be used to classify soft spots relaxing fast against slowly-relaxing regions in glasses. Here, soft spots are defined as regions that had a high likelihood of rearranging within a short time scale. As input to this algorithm, each particle was assigned a vector of local structural descriptors that captured the local density and angular structure within shells at different distances from its center. 
From the trained SVC a continuous structural descriptor called softness, $S$, can be extracted which correlates with the likelihood for the particle to rearrange in the near future. Softness has been used to gain insight into a variety of glass problems encompassing many different types of glassy liquids and disordered solids, ranging from strong to fragile and ductile to brittle with constituent particles ranging from atomic to granular, studied in bulk and in thin films
~\cite{schoenholz2016structural,SchoenholzPNAS2017,SussmanPNAS2017,CubukScience2017,HarringtonPRE2019,MaPRL2019,Cubuk2020,RidoutPNAS2022,TahJCP2022,Liu2022deciphering}. Further, this approach also led to a series of papers~\cite{ZhangPRX2021,ZhangPRR2022,xiao2023machine} aiming at the construction of an effective model for the dynamics in glassy fluids built around the evolution of the softness field, which will be discussed in more detail in Sec.~\ref{sec:models}. 

Whereas softness is associated to prediction on short time scales, several subsequent works have focused on predicting the dynamics, or more precisely the dynamic  propensity~\cite{widmer2004reproducible}, on longer time scales $t \sim \tau_\alpha$. Here, $\tau_\alpha$ is the typical structural relaxation time scale on which each particle moves on average roughly one particle diameter (see SM \cite{SM}). Dynamic propensity quantifies the local dynamics of a glass-forming liquid by capturing the typical behaviour of each particle in a structure using the isoconfigurational ensemble~\cite{widmer2004reproducible}. This ensemble is formed by a set of trajectories which start from the same initial structure but with different initial velocities drawn from a Maxwell-Boltzmann distribution.  By calculating the mean distance 
travelled by a given particle in this ensemble, one arrives at the dynamic propensity $\langle |\Delta \mathbf{r}_i(t) |\rangle_\text{iso}$. Due to this averaging, the dynamic propensity captures the part of the dynamics encoded in the initial structure, leaving out the part stemming from the initial velocities which cannot be captured by any structural descriptor. While some dynamical information is necessarily lost~\cite{berthier2007iso}, the dynamic propensity, which fluctuates from one particle to the other, is an important measure for dynamic heterogeneity in supercooled liquids. At the end of the section, we will also discuss ways to reintroduce the fluctuations around the isoconfigurational average using different labels or new ML designs.

In 2020, Bapst \textit{et al.} \cite{bapst2020unveiling} introduced a graph neural network (GNN) that could predict the dynamic propensity significantly better than a support vector machine (SVM) (where the input for the SVM was angular and radial functions similar to Ref. \cite{CubukPRL2015}).  In contrast to the SVM, the input to the GNN was a graph structure where each particle in the initial configuration was represented by a vertex, and edges were drawn between particles within a cutoff radius of each other. In addition to the structure of this graph, the vertices and edges also carry information: the particle species (as vertex data) and the vectors connecting neighbours (as edge data). The GNN model then consists of several multilayer perceptrons (MLP) that iteratively update the features contained at the edges and nodes, with each iteration passing information along the nodes and edges of the network. After the final iteration, the features at the nodes are passed through a final MLP that predicts the dynamic propensity.
As all vertices are updated in parallel, the network predicts mobilities of all particles simultaneously. This network was then optimized to minimize the squared difference between the predicted and true propensity using the $\mathcal{L}_2$ norm. Recently, two improvements to this GNN approach were proposed.  Shiba \textit{et al.} \cite{GNNrelative2022} demonstrated that a significantly higher accuracy could be reached over nearly all time scales by not only considering single particle dynamics but also pairwise dynamics. Specifically, they trained the GNN to predict not only the dynamic propensity but also the isoconfigurational change in the distance between pairs of particles sharing an edge in the graph --- a modification called BOnd TArgeting Network (BOTAN)~\cite{GNNrelative2022}. Intriguingly, even with the same overall architecture, BOTAN finds a better prediction for the single particle dynamics, showing that the extra edge information improves the performance of the GNN. Pezzicoli \textit{et al.} explored an alternative improvement by explicitly requiring the GNN to enforce the rotational symmetry, an idea sometimes referred to as geometric deep learning or rotation equivariant network (SE(3), \cite{pezzicoli2022se}).  This adaptation also improved on the original work of Ref.~\cite{bapst2020unveiling} over nearly all considered time scales.

In addition to the development of increasingly sophisticated ML methods to predict the dynamic propensity, significant efforts have been made to better capture important local features of the structural input.  Shortly after the seminal work on GNNs \cite{bapst2020unveiling}, Boattini \textit{et al.} incorporated the recursive updating properties of GNNs into a set of locally coarse-grained structural descriptors. They found that fitting just three generations of descriptors with a linear regression algorithm was sufficient to essentially reach the accuracy of the GNN --- leading to a much simpler and more interpretable algorithm for fitting glassy dynamics. Interestingly, learning the dynamic propensity using these descriptors with non-linear models (MLP and GNN) did not improve the ability to predict the dynamics, as quantified by the Pearson correlation coefficient~\cite{Filion2022}.  

This approach has been further developed by including physics-inspired descriptors which have been identified as important structural proxies in the past thirty years of glass research (GlassMLP, \cite{jung2022predicting}). These additional structural descriptors include potential energy and properties of the Voronoi cells, as well as the choice of describing the system in terms of its inherent state~\cite{jung2022predicting}. The inherent state corresponds to the energy minimum configuration that is closest to the actual input structure. These modifications improve the performance of the network. GlassMLP further uses MLP for supervised learning, which enables the precise representation of non-linear or non-Gaussian features such as probability distributions of propensities~\cite{jung2022predicting}. Using transferability in system size, the network has been applied to determine dynamic correlation lengths and the geometry of rearranging clusters over a wide range of temperatures. The GlassMLP model was recently enhanced to improve the transferability across time scales and temperatures and to explore physical regimes where direct training cannot be performed~\cite{jung2023dynamic}. In a vein similar to GlassMLP, structural descriptors have been improved by going beyond inherent states and using cage states (CAGE, \cite{alkemade2023improving}). Cage states are extracted from restricted ensemble averages of the local structure using Monte Carlo simulations and thus better describe the local environment, which also helps improving the performance of the model.

By construction, all methods perform best at the temperature at which they are trained. 
However, it is possible to apply a trained network to other temperatures, and test how predictions correlate with true dynamics there~\cite{CubukPRL2015,bapst2020unveiling,TahJCP2022,pezzicoli2022se,jung2023dynamic}. Good performance in such transfer experiments indicates that the model captures relevant universal features in the structure-dynamics relationship of glass-forming liquids. Transferability has, for example, been used in Ref.~\cite{TahJCP2022} to investigate links between amorphous structure and fragility and in Ref.~\cite{jung2023dynamic} to predict features of dynamic heterogeneity for temperatures comparable to the experimental glass transition temperature. In Sec.~\ref{sec:benchmark}, we show additional transferability experiments for several models. 

In future work, it will be interesting to further exploit the transferability of supervised ML techniques to robustly extract information on structural relaxation at very low temperatures. One particular goal would concern the evolution of dynamic heterogeneity upon approaching the glass transition, and make the connection with experimental results~\cite{bookDH}. One possible strategy to improve transferability of trained models to low-temperature regimes in which dynamics cannot be run for long enough (i.e., no or little labeled data is available) would be to use self-supervised learning~\cite{gidaris2018unsupervised}. More generally, self-supervised learning could also be used to enhance performance of the deep approaches (see also Sec.~\ref{sec:supervised_RL}).

To better capture the physical phenomena underpinning glassy dynamics, it seems necessary that the ML models not only faithfully predict the dynamics at the single particle level, but also correctly reproduce all statistical features of the propensity field, including spatial and temporal correlations, such as those measured by the four-point susceptibility $\chi_4(t)$ \cite{PhysRevE.71.041505}. For GNNs, a known problem is over-smoothing, i.e., the predicted propensity field tends to be smoother than the ground truth. Jung \textit{et al.}~\cite{jung2022predicting} introduced additional correlations-related terms in the loss function to avoid over-smoothing, and could show that the predictions indeed display more realistic correlation functions, even in a rather simple MLP architecture. An open direction for future works is the development of such improved loss functions, which could also be used in deep architectures.

A related idea is to learn not only the isoconfigurational average of the displacement, $\langle \Delta \mathbf{r}_i(t) \rangle_\text{iso}$, but its full statistical distribution, $P_\text{iso}(\Delta \mathbf{r}_i(t))$. This could be accomplished either by additionally fitting higher moments of the distribution, or by taking a generative model approach, in which one would attempt to generate realistic single-instance dynamical fields (not averaged in the isoconfigurational ensemble). This task could be achieved using a variational auto-encoder approach~\cite{kingma2019introduction}. The spatial correlations of the isoconfigurational average being a priori different from the  spatial correlations of single instances, the hope is that such generative models would reproduce these statistics more faithfully than conventional ones. To train such models, one should probably use single-instance configurations, which can also be used to regress the average. Finally, this approach might then be usable to propose new configurations on the structural relaxation time scale $\tau_\alpha$, with the long-term goal to be able to develop a Monte Carlo algorithm which completely avoids the critical slowing down of the dynamics on the approach to the glass transition. Similarly, the predictions could be used to create ultra-stable glasses by generating prototypical hard neighborhoods and remove structural ``defects''~\cite{wang2021inverse}. In Sec.~\ref{ref:generative} we discuss in more detail generative models and their applications for sampling low-temperature glassy structures.

In physics, explaining complex behavior builds on the ability of theories and models to substantially compress the inherent information of a natural phenomenon~\cite{Kivelson_Kivelson_2018}.
From this standpoint, large machine learning models, involving several hundreds or thousands of directly fitted parameters, do not qualify as a physical theory in the traditional sense of the term.
Moreover, due to their non-linearity, neural-networks models are still rather difficult to interpret, although some progress is being made~\cite{wang2020predicting, miao2022interpretable}.
This does not mean, however, that large ``black-box'' ML models are useless in this context: their predictions can be instrumental as part of an euristic process, eventually leading to a simple solution to an outstanding problem~\cite{Duede_2023}.
This is nicely demonstrated by the results presented in this section: building on the insight of Ref.~\cite{bapst2020unveiling}, Filion and coworkers achieved about the same prediction accuracy as graph neural networks using a much simpler and transparent linear regression method~\cite{Filion2022}.
Linear models thus retain a strong appeal for fundamental research in glass physics, because of their simplicity and their direct mapping to the underlying structural descriptor.
Interpreting the outcome of machine learning models also hinges on the ability of identifying the most relevant features -- a process known as ``feature selection''.
Analysis of the so-called ``information imbalance'' has emerged as a general and elegant approach to feature selection~\cite{Glielmo_Zeni_Cheng_Csanyi_Laio_2022}.
This method has been very recently applied to identify the most relevant structural features for glassy dynamics~\cite{Sharma_Liu_Ozawa_2024}.

Finally, a natural aim for future investigations is to enlarge the ML studies described above to encompass diverse glass-forming materials with complex dynamics, including active glasses as model for biological tissues~\cite{berthier2019glassy,PhysRevResearch.6.023257}. Although it is generally accepted that equilibrium microscopic dynamics do not influence long-time structural relaxation it is unclear which signature activity plays on structural descriptors in active glasses~\cite{janzen2023dead}. Similarly, very little information exists on the dynamical properties of glasses during aging \cite{PhysRevMaterials.8.025602}, and whether similarly strong structure-dynamics relationships can be found as for equilibrium relaxation.  

\section{Machine Learning Phenomenological Glass Models}

\label{sec:models}

In this section, we demonstrate the potential of combining ML methods with physical insights to develop effective models and phenomenological theories of slow and glassy dynamics. The starting point are the ML methods, already described in Sec.~\ref{sec:dynamics}, that identify the local structure responsible for local relaxations on short time scales. The aim is now to use this structural field as a building block to construct a phenomenological model for how structural relaxation proceeds.  This approach is built upon two useful classes of models to understand dynamics in glass-forming liquids and amorphous solids subjected to mechanical strain, namely trap and elasto-plastic models.

Trap models start from the high-temperature liquid, describing the system as a distribution of energy barriers for rearrangements. 
A recent implementation~\cite{scalliet2021excess} adds a facilitation mechanism: in response to a rearrangement, all the energy barriers are subject to a small random drift. This facilitated trap model has been used to explain the emergence of excess wings in the relaxation spectra~\cite{guiselin2022microscopic}. Elasto-plastic models start instead from the low-temperature solid, describing the system in terms of thermally-induced rearrangement events that can trigger other rearrangements via long-ranged strain fields~\cite{RevModPhys.90.045006,ozawa2022elasticity,tahaei2023scaling}. Both of these approaches have given insights into glassy dynamics, and include facilitation effects in some manner, but both describe the local structure of the liquid in a very coarse-grained, simplistic manner. As a result, both classes of models must make ad hoc assumptions on the nature of the local relaxation events. In the case of the trap model, facilitation is assumed to lead to a shift of energy barriers, while in elasto-plastic models the distribution of yield stress is imposed. 

A recent approach unites the trap and elasto-plastic models by extending them to include the local structure in the form of a machine-learned microscopic structural descriptor, the softness $S$, as introduced in Sec.~\ref{sec:dynamics}. Softness is sufficiently accurate that the probability of rearranging for particles of a given softness $S$, $P_R(S)$, has an Arrhenius temperature dependence: $P(R|S) = \exp\left[ - \Delta E(S) / T + \Delta \Sigma(S)\right]$, where $\Delta E$ is the energy barrier and $\Delta \Sigma$ is the entropic one, $\Delta F =\Delta E - T\Delta \Sigma$. This suggests that particles of softness $S$ have a well-defined free energy barrier for rearrangements, $\Delta F(S)$. The spatial variation of the softness then leads to a free-energy barrier field that couples to the stress and strain fields. Note that any local structural measure that predicts rearrangements or local yield stress~\cite{lerbinger2021relevance} (whether extracted using ML, as described here or in other sections of this paper, or by other means~\cite{richard2020review}) could in principle be exploited in a similar manner to construct phenomenological models of glassy dynamics or plasticity.  

Thanks to this ability to extract a free energy barrier estimate $\Delta F(S)$, the softness can naturally be used to construct a phenomenological trap model. However, softness allows one to go further by considering spatial correlations. When a rearrangement occurs, it alters the softness of rearranging particles as well as that of nearby particles through near-field facilitation~\cite{ChackoPRL2021}.
It also alters the softness of more distant particles by creating a strain field that decays away from the rearrangement. Because the strain field changes the local structural environment of particles, it alters their softness. This far-field form of facilitation is well-captured by elasto-plastic models \cite{ozawa2022elasticity}.

There is an interplay between rearrangements (strain), changes in softness and elasticity, with each one affecting the other two. A systematic approach to disentangling all of these effects has been introduced in Ref.~\cite{ZhangPRX2021} and implemented in a lattice structuro-elasto-plasticity model~\cite{ZhangPRR2022} for athermal systems under load. It has been applied successfully to a number of systems of varying ductility~\cite{ZhangPRR2022,xiao2023machine} and used to extract insights into the microscopic factors that control ductility, such as the strength of near-field facilitation~\cite{xiao2023machine}. 

These results pave the way towards models of structural relaxation dynamics in glassy liquids. A simple trap model built upon the barriers $\Delta F(S)$ and assuming an underlying distribution of softness, $\rho(S)$, as in Ref.~\cite{BouchaudJPA1996}, was developed in Ref.~\cite{RidoutEPL2023}. One can also construct a version of the facilitated trap model of Ref.~\cite{scalliet2021excess} that incorporates $S$. Generalising such models to supercooled liquids, however, is more challenging since one must include time-reversal invariance. Above the mode-coupling temperature, $T_c,$ near-field facilitation should be sufficient. The hypotheses that $\rho(S)$ is nearly Gaussian~\cite{schoenholz2016structural} and that the near-field distribution of the change in softness $\Delta S(r)$ due to a rearrangement at the origin is also nearly Gaussian, are important simplifications that allows formulation of a closed theory~\cite{ridoutArXiv2024}. For systems below $T_c$, however, it has been shown that long-ranged facilitation via strain occurs~\cite{ChackoPRL2021}. The inclusion of time-reversal invariance in a model such as a thermal elasto-plastic model~\cite{ozawa2022elasticity} or structuro-elasto-plasticity model with long-ranged facilitation is a challenging open problem that needs to be solved. A precise predictor of future dynamics, as discussed in Sec.~\ref{sec:dynamics}, would be a very useful tool to adjust such models. {Additionally, it would be interesting to explore whether the free energy barriers $\Delta F$ for local relaxation, extracted from the amorphous structure using softness, could be learned more directly and used to improve effective glasss models. }

Another path forward is to switch from a field picture to a defect picture. Most predictors of rearrangements yield particle-based quantities that are easily converted to fields, but highlight localised regions that are susceptible to rearrangement~\cite{richard2020review}. These regions can be viewed as structural defects that interact with each other and are created and destroyed by strain and rearrangements. ML could be used to learn these interactions and rules, to help build defect theories of plasticity and glassy dynamics.

In this respect, a direction that has been recently explored consists in letting a glass-forming liquid evolve via the usual thermal motion (e.g., using molecular dynamics). During such exploration, the energy is periodically minimized in order to extract a library of mechanically stable zero-temperature configurations (inherent structures). The idea is then to use ML to classify pairs of inherent structures, instead of single configurations, and check whether the pair is connected by a low-energy excitation corresponding to a localised structural defect. For example, in Ref.~\cite{ciarella2023finding}, a large library of inherent structures was constructed by such an exploration at very low temperature inside a glass basin. By means of supervised learning techniques, it was possible to train a machine that inputs a pair of inherent structures and outputs, with good precision, the classical energy barrier separating them. This allowed for a speed up in the search for glass defects by more than one order of magnitude, which is significant given the complexity of these kind of calculations. 

While this preliminary study was mostly focused on very low-energy defects that are associated with thermodynamic and transport anomalies of the glass at cryogenic temperatures~\cite{ciarella2023finding},
it should in principle be straightforward to extend the techniques to detect other kinds of defects, such as those associated with plastic events under shear~\cite{richard2020review, richard2023detecting} of relaxation events under equilibrium thermal motion~\cite{Berthier2022}. This is a promising direction for further research. 

\section{Performance metrics and benchmarking}

\label{sec:benchmark}

\begin{table*}
    \centering
    \begin{tabular}{c|l|c|c|l|l|c|c}
          & Training &  ML approach &  free parameters  &states&  training time& training hardware &app. time\\
          \hline
         BOTAN \cite{GNNrelative2022} &supervised &  GNN& {54200} &th.& $\sim \text{hours}$ & {NVIDIA A100} & $\sim s$ \\
         \hline
         CAGE \cite{alkemade2023improving} &supervised &  Ridge regression&  2775  &th.+cage&  $\sim \text{min}$& CPU & $\sim \text{hours}$\\
         \hline
         GlassMLP \cite{jung2022predicting} &supervised &  MLP&    615&th.+inh.&  $\sim \text{min}$& CPU & $\sim s$\\
         \hline
         SE(3)  \cite{pezzicoli2022se}&supervised &  GNN&  52660  &th.+inh.&  $\sim \text{hours}$& NVIDIA Tesla V100 & $\sim s$\\
         \hline
 SBO \cite{coslovichDimensionalityReductionLocal2022} &unsupervised & PCA& 0  &th.&  -- & CPU & $\sim s$\\
    \end{tabular}
    \caption{Overview of the different techniques benchmarked in this roadmap. The column ``states'' refers to the usage of thermal (th.), inherent (inh.) or cage states. The training time corresponds to the time required to train one model at a specific temperature and time for the KA system in three dimensions. The application time corresponds to the time required to calculate the prediction for a single configuration, including preparation such as calculation of inh./cage states.}
    \label{tab:overview}
\end{table*}

An essential ingredient to fuel further development of ML techniques are detailed benchmarks for existing datasets, which may allow every researcher to independently develop and test new ML approaches without complex production and preprocessing of data~\cite{paperswithcode}. We provide such benchmarks for ML glass-forming liquids for different systems, different dynamical observables, and different metrics. 

The benchmarks are based on the dataset ``GlassBench'' that we have created and made publicly available (\href{https://doi.org/10.5281/zenodo.10118191}{https://doi.org/10.5281/zenodo.10118191}). The whole dataset is separated into a training set which, as the name suggests, can be used to train the neural network, and a test set which should be used only for benchmarking. In addition to initial amorphous structures and trajectories, we provide precalculated dynamical descriptors and propensities, as introduced in Sec.~\ref{sec:dynamics}. We have also uploaded sample python code for reading and processing. Additional technical information on the data format is provided with the dataset.  

The tasks identified for ``GlassBench" are directly related to the open questions highlighted in the introduction: (A) Train a model to predict single particle propensity purely from structural properties. Accuracy will be quantified using the Pearson correlation coefficient. Higher accuracy in the prediction indicates that the learned structural descriptor is indeed an important precursor for future relaxation but it can also become essential when using the model to generate new configurations. (B) Train a transferable model such that it can be accurately applied to different temperatures. This is an important task to enable investigation of structural relaxation at  temperatures that are unreachable for numerical simulations. (C) Train a model which correctly predicts spatial dynamic heterogeneity (DH) as quantified by the dynamic susceptibility. The length scale of DH is growing with decreasing temperature such that at very low temperatures some regions in the system actively rearrange while others are completely frozen.  DH are not only important for properties of glass-forming materials but are also core to fundamental theories of the glass transition \cite{ediger2000_dh,bookDH}.  

The ML techniques used for the benchmarking are summarized in Tab.~\ref{tab:overview}. In the following, we will refer to them just as models. A large variety of different models is represented, with very different numbers of fitting parameters and training time. Furthermore, the models use various ways to physically preprocess the structural input, either by using inherent states \cite{jung2022predicting}, or even by performing a Monte Carlo averaging of local cages \cite{alkemade2023improving}. These different factors, combined with the benchmarking provided below, should help choosing the most suitable method for a given purpose, with focus on either the highest-scoring predictions, computational efficiency, or interpretability. In addition to these ML techniques, we also include the performance of traditional structural descriptors based on physical intuition~\cite{tanaka2019revealing,tong2019structural,jung2022predicting}. The model of glass-forming liquid that we selected to develop GlassBench is the very popular Kob-Andersen (KA) mixture, that we study both in two and in three dimensions. 

\begin{figure}
    \centering
    \includegraphics[width=0.99\linewidth]{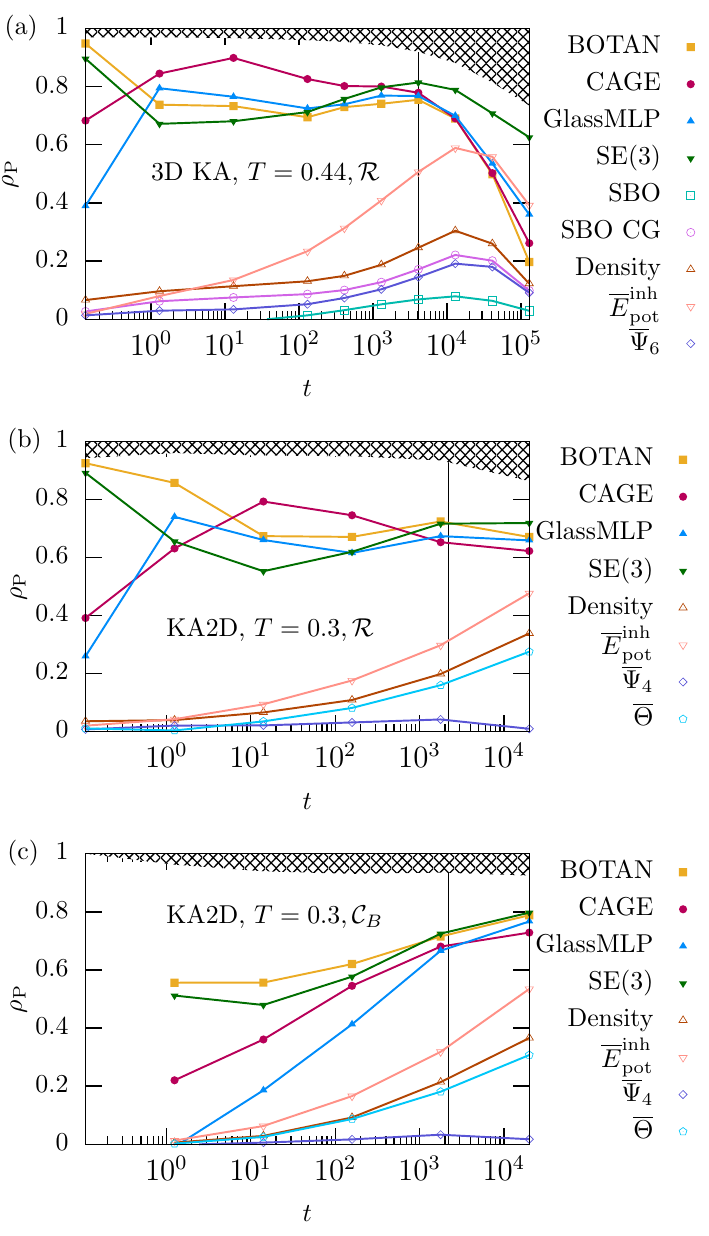}
    \caption{{\textbf{Benchmark task (A)} Pearson correlation $\rho_\text{P}$ between various structural indicators with the ground truth. (a) 3D KA model at $T=0.44$, (b,c) KA2D system at $T=0.3$. The dynamical variable is (a,b) the propensity of displacements $\mathcal{R}$ and (c) the bond-breaking propensity $\mathcal{C}_B$. The vertical line marks the structural relaxation time $\tau_\alpha$, the typical time scale on which particles rearrange, as defined in Eq.\,(1) in the SM \cite{SM}. The exclusion zone on the top marks the highest achievable correlation given the finite number of replicas.}
    }
    \label{fig:KA_pearson}
\end{figure}

{\bf Task A}. The aim for the models is to learn correlations between the amorphous structure and the dynamic propensity of displacements, $\mathcal{R}_i(t) \equiv \langle \Delta \mathbf{r}_i(t) \rangle_\text{iso}$, as introduced in Sec.~\ref{sec:dynamics}. A common metric used to assess the performance of different techniques is the Pearson correlation coefficient, 
\begin{equation}
\rho_P = \frac{\text{cov}(\mathcal{R}^\text{MD}_i, \mathcal{X}_i^\text{ML}) }{ \sqrt{\text{var}(\mathcal{R}^\text{MD}_i) \text{var} (\mathcal{X}_i^\text{ML})}},
\end{equation}
between the labels $\mathcal{R}_i^\text{MD}$ for each particle $i$ of type 1 in the entire dataset as obtained from molecular dynamics (MD) simulations, and the ML output $\mathcal{X}_i^\text{ML}$~\footnote{Calculating the Pearson correlation for each structure individually and then averaging yields slightly different results, and Pearson correlations appear to be systematically higher. Similarly, calculating Pearson correlation over particles of different type significantly increases the correlation and this should be avoided.}. The results are shown in Fig.~\ref{fig:KA_pearson}a, with full symbols corresponding to supervised ML techniques (introduced in Sec.~\ref{sec:dynamics}) and open symbols to unsupervised techniques or physically-motivated structural descriptors (Sec.~\ref{sec:structure}). All technical details are provided in the SM \cite{SM}. From this figure, it can be concluded that supervised techniques nearly approach the maximal achievable correlation over the entire range of time scales. CAGE performs best for shorter times, which appears reasonable as it uses extensive Monte-Carlo simulations to characterize the local cage structure at short times. For longer times, the SE(3) GNN extension has the strongest correlation, closely trailed by the other advanced techniques. We can also observe that there is a pronounced gap between the supervised and the unsupervised techniques, indicating that the amorphous structural features that are predictive for dynamics do not stand out in a purely structural analysis, as discussed in Sec.~\ref{sec:structure}.

To asses the generality of these findings, we also provide benchmarks for a two-dimensional ternary mixture of Lennard-Jones particles (KA2D) in Fig.~\ref{fig:KA_pearson}b. We find that, apart from minor differences, the performance of the individual techniques is very similar to the 3D system. The most noteworthy difference is perhaps that BOTAN performs best of all methods at the structural relaxation time $t\approx\tau_\alpha$ while in the 3D KA system, the trend is reversed.  While there might be subtle differences between structural relaxation across spatial dimensions the above observation implies that the problem of learning correlations between structure and dynamics is essentially independent of the spatial dimension.

The propensity of displacements $\mathcal{R}_i$ is only one specific choice to characterise relaxation dynamics among many others. One alternative is to use the bond-breaking correlation, $\mathcal{C}_B^i$, which quantifies how many nearest neighbours particle $i$ has lost during the relaxation process~\cite{Berthier2022,jung2022predicting}. As shown in Fig.~\ref{fig:KA_pearson}c, the models also successfully learn correlations between the bond-breaking propensity and the amorphous structure. The performance of the models in the short-time predictions, however, is significantly reduced compared to the propensity of displacements. This implies that predicting the exact nature and position of the first rearrangement events appears more difficult than simply predicting short-time displacements. Around the structural relaxation time $\tau_\alpha$ and beyond, the correlations shown in Fig.~\ref{fig:KA_pearson}c for $\mathcal{C}^i_B$ are stronger than for $\mathcal{R}_i$ in Fig.~\ref{fig:KA_pearson}b. This seemingly surprising result is connected to the growing dynamic heterogeneity at longer times (see also Fig.~\ref{fig:KA_chi4}), which simplifies the prediction of larger rearranging clusters from coarse-grained structural properties, as discussed in Ref.~\cite{jung2022predicting}. Additionally, at times $t \geq \tau_\alpha$, we observe that the propensity of displacement $\mathcal{R}_i$ has slowly decaying tails which are not captured by the models (see Fig.~S2 in the SM) and thus likely reduce the Pearson correlation.

\begin{figure}
    \centering
    \includegraphics[width=\linewidth]{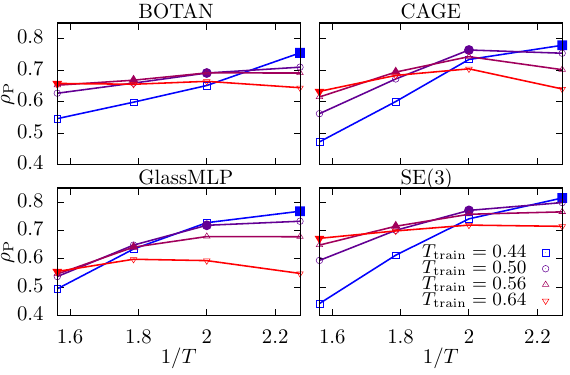}
    \caption{\textbf{Benchmark task (B)} Transferability in temperature $T$ of trained networks in the KA system (3D) at the structural relaxation time $\tau_\alpha$. Each network is trained at $T_\text{train}$ (marked by the full symbol) and applied to all four different temperatures (marked by open symbols). The color codes for $T_\text{train}$ which changes smoothly from red ($T=0.64$, high temperature) to blue ($T=0.44$, low temperature).}
    \label{fig:KA_transferability}
\end{figure}

\begin{figure}
    \centering
    \includegraphics[width=\linewidth]{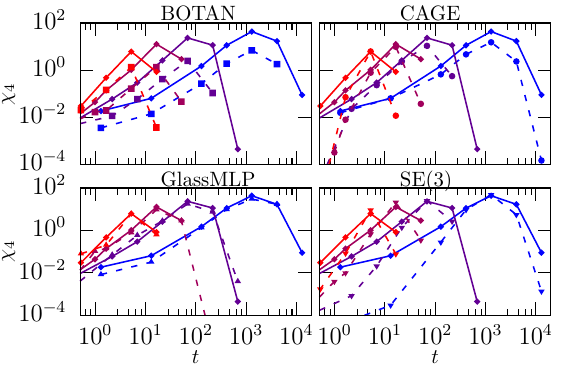}
    \caption{\textbf{Benchmark task (C)} Dynamic susceptibility $\chi_4(t)$ for different temperatures ($T=0.44$, $0.5$, $0.56$, $0.64$), as predicted from ML techniques (dashed lines), compared to the ground truth (full lines) in the KA system (3D). The color code is as in Fig.~\ref{fig:KA_transferability}.}
    \label{fig:KA_chi4}
\end{figure}

{\bf Task B.} An important property of supervised ML techniques is their transferability, in particular towards lower temperatures, with the goal to predict the dynamics at very low temperatures that are inaccessible by direct computer simulations \cite{pezzicoli2022se,jung2023dynamic}. In Fig.~\ref{fig:KA_transferability}, we evaluate the capabilities of the models to transfer their structure-dynamics relationships learnt at a given temperature to make predictions at a different temperature. The results are actually quite remarkable as transferability is generically quite good for all models. This shows that these relationships only evolve quite smoothly across the range of temperatures investigated here. In particular, the models trained at $1/T=2.0$ ($\tau_\alpha = 210$) perform nearly as good in predicting propensity at $1/T=2.25$ ($\tau_\alpha = 4100$) as the models trained directly on $1/T=2.25$. The SE(3) method seems to be particularly  suited to transfer to lower temperatures, opening the possibility to study structural relaxation at much lower temperatures.

{\bf Task C.} Finally, we investigate the performance of the models to predict the correct extent of dynamic heterogeneity. A time-dependent scalar that quantifies heterogeneities is the dynamic susceptibility 
\begin{equation}
\chi_4(t)= N \left( \langle \bar{C}_\mathcal{R}^2(t)  \rangle - \langle \bar{C}_\mathcal{R}(t)  \rangle^2  \right) 
\end{equation}
calculated from the system-averaged overlap function $\bar{C}_\mathcal{R}(t) = (1/N) \sum_{i \in N}\Theta\left( 0.3 - \mathcal{R}_i(t) \right)$. Here, $N$ is the number of particles in the system, and $\Theta(x)$ the Heaviside function. This definition separates particles into active ($\mathcal{R}_i > 0.3$) and frozen ($\mathcal{R}_i \leq 0.3$). The threshold value of 0.3 is a common choice~\cite{glotzer2003,Flenner2010} and corresponds to values slightly larger than the plateau in the mean-squared displacement~\cite{glotzer2003}, implying that particles identified as active have typically left their initial cages. In Fig.~\ref{fig:KA_chi4}, we compare the results of the predictions to the ground truth MD simulations. Despite the overall good performance in the Pearson correlation, we now observe strong differences between the various techniques. For example, the improvement in performance of SE(3) compared to BOTAN shown in Fig.~\ref{fig:KA_pearson} can be connected to their different learning of the correct dynamic heterogeneity. In a recent attention-based GNN extension, this heterogeneity has been explicitly targeted during the training procedure to improve the performance of the deep network~\cite{jiang2022geometry}. The best overall performance in predicting $\chi_4(t)$ is achieved by GlassMLP, which was specifically constructed to learn and predict dynamic heterogeneity~\cite{jung2022predicting}. This analysis shows that the Pearson correlation is not entirely sufficient to quantify the performance of a model. Additional dynamical observables, such as the dynamic susceptibility $\chi_4(t)$, should be investigated to better characterize the ability of models to realistically describe structural relaxation in glass-forming liquids.

In the SM \cite{SM} we provide further benchmarking, by investigating: (a) the coefficient of determination $R^2$, another popular measure to study the performance of ML models; (b) the probability distributions of the predicted propensities, which correspond to a second contribution to dynamic heterogeneity; (c,d) the scatter plots and the snapshots directly comparing the true and predicted propensities; (e) the cross-correlations between the different structural descriptors and the models investigated in this section. We also provide additional information on (f) learning curves, (g) transferability to lower temperatures $T=0.4$, and (h) bond-breaking propensity. 

Another conclusion of additional benchmarks, which goes beyond the scope of this roadmap, is that correlation coefficients display some system dependence~\cite{hocky2014correlation}. Hard sphere glasses and systems with strong icosahedral order, for example, show systematically higher correlation coefficients between structural and dynamical descriptors than Kob-Andersen mixtures~\cite{Filion2022, coslovichDimensionalityReductionLocal2022}. The same statement holds true for different dynamical descriptors and coarse-grained quantities. We therefore strongly encourage the use of identical datasets and labels (i.e., dynamical descriptors) to enable comparability. 

Among possible extensions of GlassBench, it would be particularly worth including a diverse set of models of glass-formers: fragile molecular glass-formers, strong network-forming glasses and potentially metallic glasses. The first step in creating a new dataset is the sampling of independent configurations at a given temperature or density. This step could leverage enhanced sampling techniques such as parallel tempering \cite{swendsen1986replica} or swap Monte Carlo \cite{berthier2016equilibrium}. Subsequently, for each structure, molecular dynamics or ab-initio dynamics simulations will have to be performed to study the structural relaxation of the systems and calculate the isoconfigurational average \cite{widmer2004reproducible}. 

It would also be interesting to broaden the tasks. For example, other ML models, such as the softness derived from SVC, target predicting local energy barriers for short-time rearrangements instead of long-time structural relaxation \cite{schoenholz2016structural,ZhangPRX2021} (see Sec.~\ref{sec:models}). For this class of ML techniques, it would be preferable that the descriptor can effectively separate the particles by their probability of rearrangement, $P_R(S)$. To encourage further development in this area, it would be desirable to perform a similar benchmarking and investigate whether modern ML models can outperform the state of the art.

Similarly, there is much interest in understanding plastic events and failure of glassy materials under external load. Contrarily to equilibrium structural relaxation, this protocol takes place far from equilibrium, but displays similar characteristics. Recently, this field has been reviewed and benchmarked in a collaborative publication~\cite{richard2020review}. The focus, however, was not yet on advanced ML techniques, which further demonstrates the very quick development of this field. It would be interesting to analyse whether the techniques presented in this roadmap can help investigating glass deformation under shear~\cite{bapst2020unveiling,fan2021predicting,du2021shear,font2022predicting}.

\section{Summary and Further Directions}

\label{sec:summary_directions}

One of the main challenges in understanding glass-forming liquids is connecting their structural properties to emergent  relaxation dynamics and plastic deformation. Since the amorphous structure lacks apparent long-range order, machine learning techniques are potentially useful to extract important information from simulations or experimental measurements and eventually generate new data. We therefore anticipate that machine learning will have a decisive influence on the field of glass physics in the upcoming years.

We have given both an overview of recent achievements in using machine learning to advance our fundamental understanding of glass physics, and discussed important research questions for the future. We identified three main fields in which machine learning could impact research on glass physics: (i) automatic feature detection to better understand locally favored amorphous structures, (ii) forecasting of future microscopic dynamics and investigation of dynamical heterogeneities, and (iii) the construction of data-driven effective models of glassy dynamics. For each of these fields we have proposed concrete methodology that could be used to answer some of the identified research questions. Finally, we have provided a dataset and benchmarking to provide a common ground to compare ML methods for glassy dynamics and enable reproducibility.  This work will hopefully provide inspiration and guidelines to researchers on how to further develop the field with the ultimate goal to better understand universal properties of glass-forming liquids and, eventually, the glass transition itself.

The number of potential applications of ML techniques for studies of glass-forming liquids and glasses goes of course much further than what we covered in this article, focusing on fundamental aspects of glassy dynamics. This includes investigation of specific material properties \cite{liu2021machine,ravinder2021artificial,bodker2022predicting,cassar_glassnet_2023} or material discovery \cite{mauro2016accelerating,tandia2019machine,merchant2023scaling}, machine learning force fields \cite{unke2021machine},
and particle identification in experimental data \cite{volpe2023roadmap,Midtvedt_DeepTrack2}. Furthermore, the connection between ML and glassy physics has also gone in the reverse direction, by borrowing methods developed for the study of disordered systems to analyze central theoretical ML problems. In fact, the connection between the rough energy landscape of amorphous materials and optimization defined by loss functions with many local minima  has been used to better understand and optimize learning of neural networks \cite{mehta2019high,gabrie2020mean,merchant2021learn2hop,gabrie2023neural,bonnaire2023high,mezard2023spin}.  

We anticipate that modern ML frameworks will continue to impact glass research and lead to the development of new major directions in the field. We close this roadmap by discussing exciting new concepts that have the potential to play an important role in future research.

\subsection{Attention and transformers}

An important advance in recent ML architectures is the ``attention mechanism"~\cite{vaswani2017attention}. Using it in ML methods for glassy dynamics has a lot of potential. The fundamental concept behind attention is to assign a learnable level of importance to specific parts of the input or intermediate representation. This could be distinct words in sentences (see, e.g., ChatGPT), amorphous structures' specific features, channels in a deep representation or neighboring atoms in a graph representation. Broadly speaking, this can be achieved by making learned weights themselves dependent on the input.

Salient examples of successful applications of attention are AlphaFold v2 and RoseTTAFold~\cite{Jumper2021, baker_rosettafold_2021}, which both use a rotation-equivariant attention-based transformer to predict the three-dimensional structure of proteins. This architecture differs from the graph network architectures discussed earlier by the dense character of its computation mechanism: all atoms in the input can exchange information with all other atoms (with a learned modulation as a function of the distance), allowing for a more flexible computation. 
Variants of these architectures tend to obtain results which are competitive with sparse graph networks on benchmark tasks \cite{Bratholm_2021}. Recently, a network with a self-attention mechanism was developed for predicting glassy dynamics \cite{jiang2022geometry}, showing that the curse of overfitting can be avoided. 
The broader concept of input-dependent weights was also used in a similar context to learn dynamic heterogeneities over a wide range of temperatures at which attention must be paid to temperature-dependent length scales \cite{jung2023dynamic}.
Both of the approaches are, however, 
rather direct applications of the idea of positional encoding for learning attention weights, and more complex networks, such as full-fledged transformers \cite{vaswani2017attention} are expected to be used in future research.

Going in the opposite direction from transformers, attention mechanisms can also be used to make models more interpretable. For example, they can help identify the task-relevant sectors in the input data~\cite{pmlr-v162-miao22a}. Additionally, attention can also be incorporated into other architectures, for example in combination with a temporal encoding in time-series forecasting \cite{qin2017dual}, to identify relevant parts of past trajectories.

\subsection{ Self-supervised, semi-supervised and reinforcement learning }

\label{sec:supervised_RL}

So far, we have mainly focused on ``traditional'' unsupervised and supervised learning techniques since they are better established in the field. However, there are several other learning paradigms which are potentially useful for future projects, some of which have started to be used in glass research.

To learn the connection between structural order and structural relaxation in a way that reconciles the unsupervised and supervised approaches, a possibility is to use semi-supervised learning~\cite{books/mit/06/CSZ2006}. Concretely, the idea is to perform self-supervised learning using only unlabeled input configurations by designing a pretext task, such as reinserting a particle which has been artificially removed from a configuration, denoising particles' positions, or predicting local quantities (such as the potential energy of each particle or its distance to its quenched position) \cite{gidaris2018unsupervised}. Once a representation has been learned to perform this mock task, one can fine-tune only a handful of parameters to correlate the learned representation to the relevant dynamical variable. 
Such self-supervised pre-training has proven effective in increasing performance for various downstream tasks that are similar in spirit to glassy dynamics prediction, such as molecular properties  \cite{rong2020self}, 
crystalline material properties \cite{magar2022crystal} 
or organic semiconductors optoelectronic properties prediction \cite{zhang2022graph}. In the application to glasses, the key element of this approach is to build most of the network without looking at labels, so the output of such a method may be more acceptable as a bona fide structural descriptor, as opposed to heavy networks relying purely on supervised learning. Additionally, it requires fewer labeled data which becomes very important at low temperatures at which sampling becomes difficult. The described methodology could therefore also be used to improve transferability of pre-trained models.

Although reinforcement learning is a well established tool in the field of machine learning \cite{kaelbling1996reinforcement} only very recently it has found applications in physics for improved sampling~\cite{shain2019sampling, fan2023searching} or structure optimization~\cite{doi:10.1021/acs.jctc.0c00971, bihani2023stridernet}. The general idea behind reinforcement learning is to learn taking specific actions when reaching certain states. The goal is to find the policy of actions leading to optimal results, as quantified by a reward function. Applying this approach to the example of searching ground states in spin glasses, the state would be the observed structure, action would be a spin-flip and reward the energy change after several spin flips \cite{fan2023searching}. Along these lines, Bojesen~\cite{Bojesen_2018} has formulated the Metropolis-Hastings algorithm in a reinforcement learning setting suitable for simulations of spin systems. Recent work by Galliano \textit{et al.} has extended this approach to learn novel Monte Carlo moves that accelerate sampling of supercooled liquids~\cite{galliano2024policyguidedmontecarlogeneral}, establishing connections with related adaptive Monte Carlo methods~\cite{Christiansen_Errica_Alesiani_2023, Gabrie_Rotskoff_Vanden-Eijnden_2022}. One of the goals of this line of research is to improve and generalize the swap Monte Carlo algorithm, which demonstrated an impressive performance for specifically adapted glass models \cite{berthier2016equilibrium,swap:ninarello2017} and has led to a series of exciting new insights into supercooled liquids~\cite{guiselin2022microscopic,Berthier2022}. Devising general-purpose enhanced sampling algorithms to simulate glassy systems represents an exciting challenge for future research~\cite{berthier2023modern}.

\subsection{Generative models}

\label{ref:generative}

Another promising emerging direction for applying ML techniques to problems relevant to fundamental aspects of glass physics is the use of generative models (GM). One of the key problems in theoretical studies of glassy dynamics is that of sampling. In fact, sampling efficiently configurations $x_i$ from a Gibbs distribution of the form $P(x_i) = \exp[-\beta U(x_i)]/Z$, where $\beta^{-1}=k_B T$ is the target inverse temperature, and $U(x_i)$ is the known potential energy is very challenging for systems exhibiting glassy dynamics. Different from generative modelling of images, in which one estimates an unknown probability distribution from data, here the target distribution is therefore known from the start and sampling from it is the challenge.

A first line of research was proposed independently by No\'e {\it et al.} under the name of Boltzmann generators~\cite{noe2019boltzmann} and by Wu {\it et al.}~\cite{wu2019solving} using variational autoregressive networks. The idea is to consider a much simpler distribution $\hat{P}(z)$, such that one can easily sample $z_i$ in a single shot. This can, for example, be a Gaussian model, an autoregressive model, or a Gibbs distribution at very high temperatures at which we can sample efficiently. After learning, independent samples $x$ can then be generated using the invertible map $x_i=f(z_i)$ of the model. By computing the biased distribution $P_\text{GM}(x)$ of the generative model enables unbiasing $x_i$ in a last post-processing step \cite{noe2019boltzmann}. This ensures that we are precisely sampling from $P(x)$.

In practice the method is only efficient if the weights are almost uniformly distributed. This requires that the generative distribution is as close as possible to the target distribution. To this aim, the machine is trained to minimize the Kullback-Leibler (KL) divergence between the generative and target distributions. Because the KL divergence between two distributions is not symmetric, two choices can be made: (i) Maximum likelihood training, in which one minimizes $D_{KL}(P || P_\text{GM})$. This approach has the advantage that $P_\text{GM}$ has to cover well all the support of $P(x)$. However, it also requires existing samples from $P(x)$, which renders training impractical: we want to train a machine to sample from the target, but for this we need to be able to sample from the target. (ii) Variational or energy-based training, where one minimizes $D_{KL}(P_\text{GM} || P)$, which corresponds to the free energy of the generative model. While this choice does not require sampling from $P_\text{GM}$ it has the important drawback that the generative model might only cover part of the support of the target, which is known as mode collapse.

To deal with these problems, several architectures and training strategies have been proposed. To sample a two-state protein model No\'e {\it et al.}~\cite{noe2019boltzmann} focused on a normalizing flow architecture to represent the map $f(z)$, and a training strategy based on mixing the variational approach with maximum likelihood, for which experimental structures and short molecular dynamics simulations were used. In subsequent work, the group introduced equivariant flows to implement physical symmetries~\cite{kohler2020equivariant}, and used a higher-temperature Boltzmann distribution as a prior~\cite{dibak2022temperature,invernizzi2022skipping} (see also Refs.~\cite{xu2022geodiff,coretti2022learning}). To sample crystalline structures, it was recently proposed to generate displacements from a reference lattice structure instead of absolute particle positions~\cite{vanleeuwen2023boltzmann}. Gabri\'e {\it et al.}~\cite{gabrie2022adaptive} introduced a more efficient training strategy based on mixing standard Monte Carlo moves with moves proposed by the generative model. Applying Boltzmann generators to sample supercooled liquids yields performances in the same magnitude as previously known enhanced sampling techniques~\cite{jung2024normalizing}.

Several studies have focused on other models in which sampling is challenging, e.g. spin-glasses, hard optimization problems and lattice field theories \cite{wu2019solving,mcnaughton2020boosting,hibat2021variational,wu2021unbiased,inack2022neural,ciarella2023machine,schuetz2022graph,schuetz2022combinatorial,fan2023searching,albergo2019flow,kanwar2020equivariant,de2021scaling,gerdes2023learning,luo2021gauge,marchand2022wavelet}. It remains unclear how these methods compare to standard ones, or whether the efficiency is universal or model dependent~\cite{angelini2023modern,boettcher2023inability,boettcher2023deep,ghio2023sampling}.

Finally, it could be worth combining the ML models discussed in Sec.~\ref{sec:dynamics}, which can precisely predict future dynamics, with generative models. The former can be used to detect active regions or even particles which are likely to rearrange, while the latter can subsequently propose new configurations based on local rearrangements.

\section*{Data availability}

The dataset ``GlassBench'' and python scripts used to create the benchmarks presented in Sec.~\ref{sec:benchmark} is publicly available and can be download on zenodo (\href{https://doi.org/10.5281/zenodo.10118191}{https://doi.org/10.5281/zenodo.10118191}).

\acknowledgments

This roadmap paper originates from discussions and interactions at the \href{https://www.cnrs.fr/fr/le-centre-artificial-intelligence-science-science-artificial-intelligence-aissai}{AISSAI} (AI for science, science for AI) workshop on ``Machine Learning Glasses'' held in November 2022 in Paris (see \footnote{Links to \href{https://www.cnrs.fr/en/workshop-machine-learning-glassy-dynamics}{Workshop} and \href{https://youtube.com/playlist?list=PLLusiyVUCDJnprBuih5yDgToIzpsIdCi4}{YouTube channel}} for details on the workshop and \href{https://youtube.com/playlist?list=PLLusiyVUCDJnprBuih5yDgToIzpsIdCi4}{YouTube} for recorded sessions). This workshop has been organized by Giulio Biroli, Ludovic Berthier and Gerhard Jung. We thank all participants for their attendance, discussions and feedback, in particular A. Banerjee, L. Janssen, M. Ruiz Garcia, S. Patinet, C. Scalliet, D. Richard, J. Rottler, O. Dauchot and O. Kukharenko for their valuable contributions. FSP is supported by a public grant overseen by the French National Research Agency (ANR) through the program UDOPIA, project funded by the ANR-20-THIA-0013-01. FSP was granted access to the HPC resources of IDRIS under the allocation 2022-AD011014066 made by GENCI.  HS acknowledges computational resources provided by ``Joint Usage/Research Center for Interdisciplinary Large-scale Information Infrastructures (JHPCN)'' and ``High Performance Computing Infrastructure (HPCI)'' in Japan (Project ID: jh230064). AJL is supported by the Simons Foundation via the Investigator Award \#327939. In addition, AJL thanks CCB at the Flatiron Institute and the Isaac Newton Institute for Mathematical Sciences under the program ``New Statistical Physics in Living Matter" (EPSRC grant EP/R014601/1), for support and hospitality. This work was supported by a grant from the Simons Foundation (\#454933, LB, \#454935 GB). GB acknowledges funding from the French government under management of Agence Nationale de la Recherche as part of the ``Investissements d’avenir'' program, reference ANR-19-P3IA-0001 (PRAIRIE 3IA Institute).\\

\bibliography{library_local}

\end{document}

\title{Supplemental material: Roadmap on machine learning glassy dynamics}

%\date{\today}

\maketitle

%%%%%%%%%% Merge with supplemental materials %%%%%%%%%%
%%%%%%%%%% Prefix a "S" to all equations, figures, tables and reset the counter %%%%%%%%%%
\setcounter{equation}{0}
\setcounter{figure}{0}
\setcounter{table}{0}
\setcounter{page}{1}
%\makeatletter
\renewcommand{\theequation}{S\arabic{equation}}
\renewcommand{\thefigure}{S\arabic{figure}}
\renewcommand{\bibnumfmt}[1]{[S#1]}
\renewcommand{\citenumfont}[1]{S#1}

In this supplemental material (SM) we provide the technical details necessary to reproduce the data shown in the benchmarking section V of the main manuscript. Additionally, we present further benchmarking.

\section{Systems and methods} \label{app:methods}

\subsection{KA system (3D)}

Most benchmarks were performed for the Kob-Andersen binary mixture (KA) in three dimensions \cite{kob1995KA}. The basic potential used in KA is the Lennard-Jones potential,
\begin{equation}\label{eq:LJ}
V_{\alpha \beta}({r}_{ij}) = \begin{cases}
4 \epsilon_{\alpha \beta} \left[\left(\frac{\sigma_{\alpha \beta}}{{r}_{ij}}\right)^{12} - \left(\frac{\sigma_{\alpha \beta}}{{r}_{ij}}\right)^6 \right] & {r}_{ij}<r^\text{cut}_{\alpha \beta}\\
0 & \text{otherwise},
\end{cases}
\nonumber
\end{equation}
where $r_{ij} = \left | \bm{r}_i - \bm{r}_j \right|$ is the distance between particles $i$ and $j$, of type $\alpha$ and $\beta$, respectively.
The KA mixture is characterized by its non-additive interactions, $\epsilon_{11}=1.0$, $\epsilon_{12}=1.5$, $\epsilon_{22}=0.5$ and $\sigma_{11}=1.0$, $\sigma_{12}=0.8$, $\sigma_{22}=0.88,$ between the two particles types. The cutoff is $r^\text{cut}_{\alpha \beta} = 2.5 \sigma_{\alpha \beta}.$

All details on the specific KA mixture and the dataset used in this roadmap are given in Ref.~\cite{GNNrelative2022}. The system, for example, includes additional contributions in the potential to make it continuous up to the second derivative. The relaxation dynamics of each structure has been simulated using $N_R=32$ independent replicas to calculate the isoconfigurational average. All results have been extracted by solely considering type 1 particles. The networks have been trained on $N_S=400$ structures, with $N_1=3277$ type 1 particles each, resulting in a total of $N_d \approx 10^6 $ data points. The testset consists of $N_S=100$ independent structures.

\subsection{KA2D system}

The two-dimensional KA2D system is a variation of the above described KA mixture. In addition to the change of dimensionality, KA2D consists of three particle types, instead of two. This change enables to apply the swap Monte Carlo algorithm \cite{swap:ninarello2017,swap:Berthier2019} and thus go to very low temperatures \cite{jung2022predicting}. All details on the KA2D system are provided in the main text and the SM of Ref.~\cite{jung2022predicting}. 

The isoconfigurational average has been calculated over $N_R=20$ independent replicas, and $N_S=1500$ structures are provided for training. Each structure consists of 600 type 1 particles, which thus also yields a total of $N_d \approx 10^6 $ data points for training.

For both the KA and the KA2D system, the structural relaxation time $\tau_\alpha$ is defined as,
\begin{equation}\label{eq:tau_alpha}
    F_s(\tau_\alpha) = e^{-1},
\end{equation}
where $F_s(t) = \langle \sum_i \cos ( q \Delta r_i(t) ) \rangle$ is the incoherent intermediate scattering function with $q \approx 2\pi / \sigma$ and $\Delta r_i(t) = \left | \bm{r}_i(t) - \bm{r}_i(0) \right|.$ The structural relaxation time $\tau_\alpha$ therefore describes the time scale on which the particles move on average approximately one particle diameter $\sigma.$

Equilibrated structures and labels calculated from the relaxation dynamics for both KA and KA2D will be uploaded in combination with this roadmap for future research (see \href{https://doi.org/10.5281/zenodo.10118191}{10.5281/zenodo.10118191}).

\subsection{Propensity (labels)}

The input of each ML method are the positions of each particle in a given structure or derived structural descriptors. The labels that are used for training and testing are dynamical quantities characterizing structural relaxation at various time scales $t.$ As introduced in Sec.~III of the main text we use propensities to remove any effects of the initial velocities. 

For each figure in the main text, except Fig.~5, the results are based on the propensities of displacement, $\mathcal{R}_i = \langle \Delta r_i(t) \rangle_{\rm iso} = N_R^{-1} \sum_{k=1}^{N_R} \left|\bm{r}^k_i(t) - \bm{r}_i(0)\right|, $ where $\bm{r}^k_i(t)$ denotes the position of particle $i$ in replica $k.$ In Fig.~5 we have reported results for the bond-breaking propensity of particle $i$, $\mathcal{C}^i_B(t) = \langle n^i_t/n^i_0 \rangle_\text{iso}$. Here,   $n^i_t$ describes the number of original nearest neighbours particle $i$ still has after a time $t$ and $n^i_0$ is the initial number of neighbours~\cite{guiselin2022microscopic}. The quantity therefore decays from $\mathcal{C}^i_B(t=0) = 1 $ to $\mathcal{C}^i_B(t\rightarrow \infty) = 0 $. Details and parameters are described in the SM of Ref.~\cite{jung2022predicting}.

\subsection{BOTAN}
The model architecture of BOTAN~\cite{GNNrelative2022} is based on that of the original GNN proposed by Bapst {\it et al.}~\cite{bapst2020unveiling}.  The difference is that BOTAN is equipped with an additional MLP layer as a decoder for targets on graph edges. BOTAN is implemented as an ``interaction network''~\cite{battaglia2018} consisting of a pair of two-layer MLPs with the size of $(64,64)$. Each of the MLPs is connected to nodes and edges, respectively, and performs message passing with the other. 

The input graphs are constructed by choosing each particle as a node of the graph, and ``connecting'' nearest-neighbour pairs by edges.  The threshold distance for the nearest neighbours are set to $2.0\sigma_{11}$ (both for KA and 2DKA). For recursive message passing, the MLPs repeat the message passing iteratively 7 (KA) or 8 (2DKA) times, but the results are not strongly affected by slight changes in this number. The target quantity on the edges is, similarly to Ref.~\cite{GNNrelative2022}, the isoconfigurational average of distance change between particle pairs which are connected via the edges. The loss function is defined as $\mathcal{L}
 = 0.4\mathcal{L}_{\mathcal{N}}  + 0.6 \mathcal{L}_{\mathcal{E}}$, where $\mathcal{L}_{\mathcal{N}}$ and $\mathcal{L}_{\mathcal{E}}$ denote the MSE losses for node target (particle propensity) and edge target (propensity for pair-distance change), respectively. Details and visualisation of the method are presented in Ref.~\cite{GNNrelative2022}.  

The BOTAN GNN model contains about 54,200 weight parameters (both for 2D and 3D).  The training is started by using pretrained model parameters (optimized to yield the edge target quantities in advance), and is performed over 1,000 epochs on 400 snapshots  (KA) and over 2,000 epochs on 1,500  snapshots (2DKA), respectively.  The batch size is fixed to one graph.  The training has been conducted separately for each temperature and time, by using the Adam optimizer with learning rates of $10^{-4}$  (KA) and $2\times 10^{-4}$ (2DKA). 
The overall training time amounted to 4 hours (KA) and 11.5 hours (2DKA) on an NVIDIA A100 Tensor Core GPU (40GB SXM) on Aquarius subsystem of Wisteria/BDEC-01 Supercomputer at Information Technology Center, University of Tokyo, with the host CPUs being 2 Intel Xeon Platinum 8360Y. These training times do not include the elapsed time of the data loader. 
The code and pretrained model parameters for 3D are provided on a GitHub repository ( \texttt{https://github.com/h3-Open-BDEC/pyg{\_}botan} ).

\subsection{CAGE} \label{sec.CAGE}
As described in Ref. \cite{alkemade2023improving}, we train a simple Ridge regression model to fit the dynamics, based on the structure of both the initial state and the cage state. The cage state is defined as the average position of particles before any rearrangement occurs. To obtain the cage state, we perform a Monte Carlo simulation where all the particles are confined to a sphere of radius $r_c^{\alpha}$.   For both glassy systems we use $r_c^{\alpha} = 1.25 \sigma_{\alpha \alpha}$  with $\sigma_{\alpha \alpha}$ the particle diameter of type $\alpha$.

To capture the local structure, we use recursive, rotationally invariant parameters that capture both the radial density, as well as the $n$-fold symmetry in various shells around each particle. For the 3D system, we use exactly the same set of descriptors as in Ref. \cite{alkemade2023improving}. 
For the 2D system we use the same density parameters as for 3D \cite{alkemade2023improving}, however the angular parameters are altered to better reflect 2D symmetry. Specifically, to capture the $n$-fold symmetry for the 2D system we use 
\begin{align*}
    \Phi^l_i(r, \delta) = \sqrt{\phi^l_i\cdot (\phi^l_i)^*}
\end{align*}
with 
\begin{align*}
\phi^l_i (r, \delta) = \frac{1}{Z}\sum_{i\neq j}e^{\frac{-(r_{ij}-r)^2}{2\delta^2}}e^{il\theta_{ij}},
\end{align*}
and
\begin{align*}
Z = \sum_{i\neq j}e^{\frac{-(r_{ij}-r)^2}{2\delta^2}},
\end{align*}
where $\theta_{ij}$ is the angle between a fixed axis (e.g., $x$- or $y$-axis) and the bond joining the $i^{th}$ particle with a particle $j$. In both 2D and 3D we consider $l\in\{1,12\}$. \\

Combining both the radial density and the angular parameters, we obtain a total of 462 parameters for the KA2D system:  
\begin{outline}
    \1 294 parameters that described the radial density up to the $5^{th}$ minimum in the pair correlation function (which is located at $4.8\sigma_{11}$). 
        \2 60 equally spaced in the interval $r/\sigma_{11}\in(0.5,2.0]$ with $\delta=0.025$.
        \2 20 equally spaced in the interval $r/\sigma_{11}\in(2.0,3.0]$ with $\delta=0.050$.
        \2 18 equally spaced in the interval $r/\sigma_{11}\in(3.0,4.8]$ with $\delta=0.100$.
    \1[] Note that the radial density functions are type specific, such that in the ternary system the total number of functions above is multiplied by three.
    \1  168 parameters that capture the $n$-fold symmetry up to the second minimum of the pair correlation function (located at $2.3\sigma_{11}$). 
        \2 14 equally spaced in the interval $r/\sigma_{11}\in[1.0, 2.3]$ with $\delta=0.1$ and $l\in[1,12]$.
    \1[] Note that the radial functions do not take the particle species into account.
\end{outline}

As described in Ref \cite{alkemade2023improving}, for the 3D system we obtain a total of 366 parameters. In addition to the zeroth order generation parameters, for both the 2D and 3D system, we additionally include two generations  of structural parameters that are iteratively averaged over the nearest neighbours (see Ref. \cite{Filion2021}).\\

The dataset on which we train our Ridge regression thus includes three generations of structural descriptors for both the initial structure, as well as the cage state. Additionally, for each particle we include the distance between the cage state and the initial positions $\Delta r_i^\mathrm{cage} = |\bm{r}_i^\mathrm{cage} - \bm{r}_i^\mathrm{init}|$. This means that the local structure of each particle in the 2D system is described by a total of 2197 parameters, and in the 3D system a total of 2773 parameters.

To predict the dynamics, we standardise the data and then train the Ridge regression model on 300 (KA2D) and 100 (3D KA)  snapshots, respectively. The only free parameter that is tuned for this ML method is the regularization parameter $\alpha$, which sets the strength of the penalty for large weights in the Ridge regression. We train the model on various values of $\alpha\in\{10^{-5}, 10^5\}$, and then choose the $\alpha$ that yields the highest correlation. The training time per timestep in both 2D and 3D is less than 10 minutes on a standard CPU.

\subsection{GlassMLP}
\label{sec:GlassMLP}

The methodology employed in this manuscript corresponds exactly to the procedure described in detail in Ref.~\cite{jung2022predicting}, including all training meta data. In particular, we calculate a set of $M_S$ physics-inspired descriptors, including the  coarse-grained local density, $\overline{\rho}^i_{L,\beta} = \sum_{j\in N^i_\beta} e^{-r_{ij}/L}$, which is coarse-grained over a distance $L$ by summing over all $N_\beta^i$ particles of type $\beta$ within distance $r_{ij} = |\bm{r}^\text{inh}_i - \bm{r}^\text{inh}_j| < 20 \sigma_{11}$ of particle $i$. Importantly, particle positions are extracted from the inherent structures $\bm{r}^\text{inh}_i$, and not directly from the thermal structure. Similarly, we include the potential energy, perimeter of the Voronoi cell and local variance of potential energy (see Ref.~\cite{jung2022predicting} for details).

Different from Ref.~\cite{jung2022predicting} we additionally include one further structural descriptor, based on the distance between the inherent and the thermal states of particle $i$, $\Delta r_i^\text{inh} = | \bm{r}_i^\text{inh}-\bm{r}_i^\text{th}|$: $\overline{\Delta r}^i_{L,\beta} = \sum_{j\in N^i_\beta} \Delta r_i^\text{inh} e^{-r_{ij}/L}.$ This descriptor does not affect the long-time prediction and was therefore discarded in Ref.~\cite{jung2022predicting}, but is reintroduced here mainly to improve the short-time performance of GlassMLP.

To predict propensity in the KA (3D) system, GlassMLP has in total 618 free parameters. In the ternary KA2D system, the number is slightly increased (765). The training just takes 2-5 minutes with an octa-core CPU or a Laptop GPU (NVIDIA T600 Laptop).

\subsection{SE(3)}

The model is exactly that presented as the main model of Ref.~\cite{pezzicoli2022se} (version 2, august 2023).

We recall the main features for completeness and to avoid possible confusion.
In the SE(3) GNN we use 8 layers. The structure of each layer is the same: $8$ channels for each $l$ component, with $l=0,1,2,3$ (namely, $l_{max}=3$).
The input data consists of the thermal positions (not quenched to the IS) and of the particles potential energy (but computed from the inherent state's particles positions).
We predict simultaneously propensities for all particle types and time steps.
The connectivity graph is computed using a threshold distance of $2.0 \sigma_{11}$.
The input node features are the particle type (one-hot encoded) concatenated with the potential energy (of the IS).
For the 2D case, we introduce an artificial third component to the position of all particles ($\forall i, z_i=0$) 
 %\gj{MISSING})
and used spherical harmonics embedding. This is overkill but is a quick way to adapt the 3D scheme to 2D data.

The radial-encoding MLP encodes the radius on (Bessel) basis functions using $10$ basis vectors, followed by a hidden layer of $16$ neurons, and uses a dropout of $0.3$. We use batch normalization between each convolution layer.
We minimise the MSE loss with a $L^2$ regularization coefficient $\beta = 10^{-7}$, using a batch size of $2$ graphs ($8$ in 2D), a learning rate of $10^{-3}$ with Adam ($\beta_1=0.99$).
As opposed to  Ref.~\cite{pezzicoli2022se}, we do not use a validation set in this manuscript and therefore do not perform early stopping. We simply take the last epoch model (the accuracy and losses are basically flat when we interrupt learning).

The model has  $52,660$ parameters for 3D data (respectively $53,394$ in 2D).
We perform $200$ epochs ($100$ in 2D), which takes approximately 11 hours (27 hours in 2D) using an NVIDIA Tesla V100 (32GB) (the CPUs on the node are 2 Intel Xeon Gold 6148 20 cores (40 threads) at 2.4 GHz (Skylake)).
Note that we train simultaneously the $10$ time steps ($6$ for KA2D) of a given state point (temperature), therefore the training does not have to be repeated for each timescale.

\subsection{SBO}

Unsupervised learning of local structure fluctuations is carried out for the 3D KA samples along the lines of Ref.~\cite{coslovichDimensionalityReductionLocal2022}, using the \texttt{partycls} Python package~\cite{Paret2021}.
Namely, we characterise the local structure around particle $i$ using the smooth bond-order (SBO) descriptor
\begin{equation}
X^\mathrm{SBO}(i) = (Q_0^S(i), \dots, Q_{l_\mathrm{max}}^S(i)) ,
\label{equ:xSBO}    
\end{equation}
where $Q_l^S$ are smoothed bond-orientational invariants of order $l$ computed over the first coordination shell of particle $i$, see Refs.~\cite{coslovichDimensionalityReductionLocal2022} for full details.
The maximum order $l_\textrm{max}$ is equal to 8.
A PCA is then carried out for each chemical species, to identify the directions in the descriptor space that capture the largest structural fluctuations.
In particular, the projection on the first principal component, $\tilde{X}_1^\mathrm{SBO}(i)$, is a measure of structural heterogeneity that is correlated to some extent with the dynamics of glassy binary mixtures~\cite{coslovichDimensionalityReductionLocal2022}.  
The correlation between the first projection of the SBO descriptor and the propensity of motion is calculated directly on the test dataset, since there is usually no separate training stage in unsupervised learning. 
Finally, along with the bare SBO descriptor, we consider a coarse-grained (CG) version
$$
\tilde{X}_i^{CG} = \frac{\sum_j \tilde{X}_j \cdot w(r_{ij};L)}{\sum_{j=1}^N w(r_{ij};L)} .
$$
where the sum over $j$ runs over the particles of the same species as $i$.
As in Ref.~\cite{tong2018revealing} we use an exponential function $w(r;L) = e^{-r/L}$, and we set $L=1$.

\subsection{Other structural descriptors}\label{sec:other_descriptors}

In addition to these models we have also included several traditional structural descriptors.

\vspace{0.2cm}

\noindent\textbf{Density:}

\noindent The density is connected to the structural input of GlassMLP (see definition Sec.~\ref{sec:GlassMLP} above). In particular, we have used $\overline{\rho}^i_{L=5,\beta=\text{all}}$, i.e. coarse-grained over a distance $L=5$ considering particles of all types. The only difference to GlassMLP is that we have evaluated the density in the thermal states.

\vspace{0.2cm}

\noindent\textbf{Potential energy $\overline{E}_\text{pot}^\text{inh}$:}

\noindent The potential energy in the inherent state, $\overline{E}_\text{pot}^\text{inh}= \overline{E}^i_{L=5,\beta=\text{all}} = \sum_{j\in N_\beta^i} E^j e^{-r_{ij}/L} / \bar{\rho}^i_{L,\beta}$, is similarly part of the structural descriptors which is used as input for GlassMLP \cite{jung2022predicting}. Here, $E^i = \sum_{j\neq i} V(r_{ij})/2$ is the potential energy of particle $i.$ 

\vspace{0.2cm}

\noindent\textbf{Bond-order 3D ($\Psi_6$):}

\noindent The bond-order descriptor $\Psi^i_6$ of particle $i$ for the KA system in three dimensions is defined via the complex coefficient \cite{steinhardt1983bond}
\begin{align}
    q_{l,m}^i = \frac{1}{N_b} \sum_{j \in N_b} Y_l^m(\bm{r}_{ij}),
\end{align}
where $N_b$ is the number of bonds of particle $i$, defined as all neighbours within a cutoff of $r_\text{cut}=2.0\sigma_{11},$ and $Y_l^m(\bm{r}_{ij})$ are the spherical harmonics. Using $q^i$ we define,
\begin{equation}
    \Psi^i_l = \sqrt{\frac{4 \pi}{2l + 1} \sum_{m=-l}^{m=l}\left| q^i_{l,m} \right|^2}.
\end{equation}
As before, we also further coarse-grain the descriptor, $\overline{\Psi}^i_{6,L=5,\beta=\text{all}} = \sum_{j\in N_\beta^i} \Psi_6^j e^{-r_{ij}/L} / \bar{\rho}^i_{L,\beta}$.\\

\vspace{0.2cm}

\noindent\textbf{Bond-order 2D ($\Psi_4$):}

\noindent For the KA2D system, we slightly adapt the definition,
\begin{align}
    q_{l}^i = \frac{1}{N_b} \sum_{j \in N_b} e^{\textrm{i} l \theta_{ij}},
\end{align}
which is strongly related to the $n$-fold symmetry defined in Sec.~\ref{sec.CAGE}. From this we calculate,
\begin{equation}
    \Psi^i_l = \sqrt{ q_l^i \cdot (q_l^i)^* },
\end{equation}
which is subsequently coarse-grained in the same way as defined in the 3D case.\\

\vspace{0.2cm}

\noindent\textbf{Tanaka's $\Theta$ order parameter:}

\noindent In Ref.~\cite{tong2019structural} Tong and Tanaka proposed the usage of a structural descriptor $\Theta^i,$ which quantifies the strength of local packing around particle $i$ (see Ref.~\cite{tong2019structural} for details and definitions). We similarly coarse-grain this descriptor $\overline{\Theta}^i_{L=5,\beta=\text{all}} = \sum_{j\in N_\beta^i} \Theta^j e^{-r_{ij}/L} / \bar{\rho}^i_{L,\beta}$.

In a 2D system of polydisperse harmonic spheres, Ref.~\cite{tong2019structural} reports a Spearman’s rank correlation coefficient of around $\rho_\text{S} \approx 0.9$ for the coarse-grained $\overline{\Theta}$ with propensity. $\rho_\text{S}$ is usually strongly related to Pearson correlation $\rho_\text{P}$. In this roadmap, we show that the performance is significant reduced when applied to systems which are not prone to crystallization. Additionally, some of the performance difference might emerge from calculating Pearson correlation over all particles independent of their radii in Ref.~\cite{tong2019structural}, as opposed to making independent predictions for each particle type as done in this manuscript.  This emphasises the importance of using standardized datasets and performance metrics to quantify and validate the performance of newly developed descriptors and techniques.

\section{Additional benchmarking} \label{app:add_benchmarking}

In the following, we will discuss additional metrics and ways to visualise the performance of the various methodologies which have been benchmarked intensively in Sec.~V of the main text.

\subsection{Coefficient of determination $R^2$}

\begin{figure}
    \centering
    \includegraphics[scale=0.92]{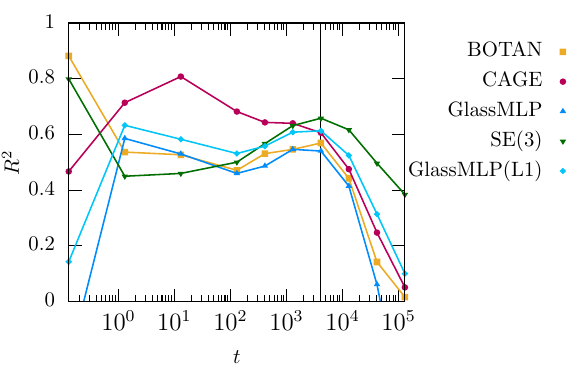}
    \caption{Coefficient of determination $R^2 = 1 - \sum_i \left( \mathcal{R}^\text{MD}_i - \mathcal{X}_i^\text{ML} \right)^2 /  \text{var}(\mathcal{R}^\text{MD}_i)$ between the model predictions and the ground truth in the KA system (3D) at $T=0.44$ for various ML models. The dynamical variable is the propensity of displacements. The vertical line marks the structural relaxation time $\tau_\alpha$. GlassMLP(L1) denotes the GlassMLP model \cite{jung2022predicting} trained without the additional terms in the loss function using only the $L^1$ norm.  }
    \label{fig:KA_R2}
\end{figure}

The coefficient of determination $R^2= 1 - \sum_i \left( \mathcal{R}^\text{MD}_i - \mathcal{X}_i^\text{ML} \right)^2 /  \text{var}(\mathcal{R}^\text{MD}_i)$ is a popular measure to quantify the quality of a fit, including, in particular, supervised ML models. In Fig.~\ref{fig:KA_R2} the coefficient of determination is shown for exactly the same data as was analysed with the Pearson correlation in Fig.~3. It can be observed that, in general, the two measures $\rho_\text{P}$ and $R^2$ do not show very different results. In particular, SE(3) still features the best performance at $t=\tau_\alpha.$ The most pronounced difference is that GlassMLP performs worse when measured with $R^2$ instead of $\rho_\text{P}.$ This can be explained by the usage of a loss function which contains additional terms on top of the standard $L^1$ or $L^2$ norm \cite{jung2022predicting}. To validate this explanation we show the results for GlassMLP(L1) trained without these additional contributions and find results significantly better than for the original GlassMLP (see Fig.~\ref{fig:KA_R2}).

\subsection{Probability distribution of propensities}

\begin{figure}
    \centering
    \includegraphics[scale=0.92]{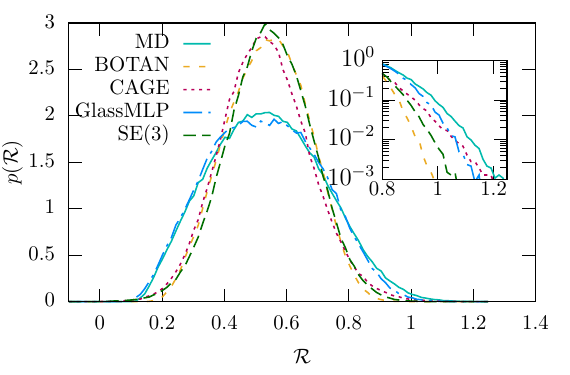}
    \caption{Probability distribution of predicted and simulated propensities at $t=\tau_\alpha$ for the KA system (3D). The dynamical variable is the propensity of displacements $\mathcal{R}.$ The inset shows a zoom to the tail, with logarithmic y-axis. }
    \label{fig:KA_histo}
\end{figure}

Another interesting observable is the probability distribution of the propensities of displacement, $p(\mathcal{R})$, which can be extracted both from the MD simulations and from the ML models. A similar analysis has been performed in Ref.~\cite{jung2022predicting}. As can be seen in Fig.~\ref{fig:KA_histo} the propensities predicted by the models tend to underestimate the variance of the distribution. This observation can be rationalized, since predicting outliers could be very costly in the loss function during training, hence networks tend to predict small variances. The only network which agrees quite well to the MD prediction is GlassMLP, which has been constructed to adapt the variance by an explicit contribution in its loss function \cite{jung2022predicting}. However, as can be seen in the inset of Fig.~\ref{fig:KA_histo}, even for GlassMLP the tail of strongly moving particles is not perfectly reproduced. Focusing on this tail would likely significantly worsen the overall performance of the model.

\subsection{Scatter plots of propensity predictions}

\begin{figure}[b]
    \centering
    \includegraphics[]{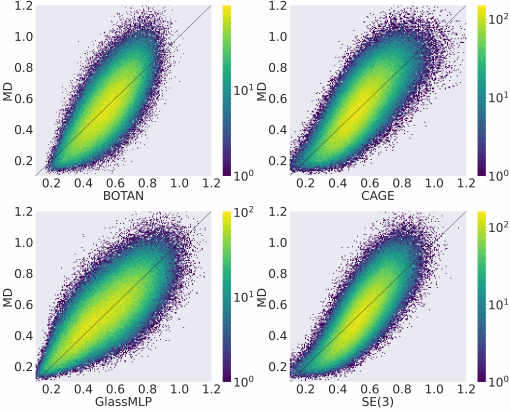}
    \caption{ 2D histograms of particle propensity $\mathcal{R}_i$ to compare different ML techniques to the MD ground truth in the KA system (3D, $t=\tau_\alpha$). Each entry in the histogram represents one particle, $i$, where the x/y-coordinates correspond to the propensities as predicted by ML and from MD simulations, respectively. }
    \label{fig:KA_scatter}
\end{figure}

The Pearson correlation $\rho_\text{P}$ or the coefficient of determination $R^2$ are very good measures to quantify the performance of a network using a scalar quantity (see Figs.~3-6 in the main text and Fig.~\ref{fig:KA_R2}). However, they obviously cannot capture all details of the connection between the true and predicted propensities. In Fig.~\ref{fig:KA_scatter} we therefore show a 2D histogram visualizing this connection in greater detail. The overall impression of these figures is very similar to what we have concluded before. The most obvious difference between the methods is that GlassMLP is symmetric around the straight line, while the other techniques are slightly asymmetric. This observation is connected to the differences in the probability distribution as discussed in the previous paragraph.

\subsection{Snapshots}

\begin{figure}
    \includegraphics[scale=0.96]{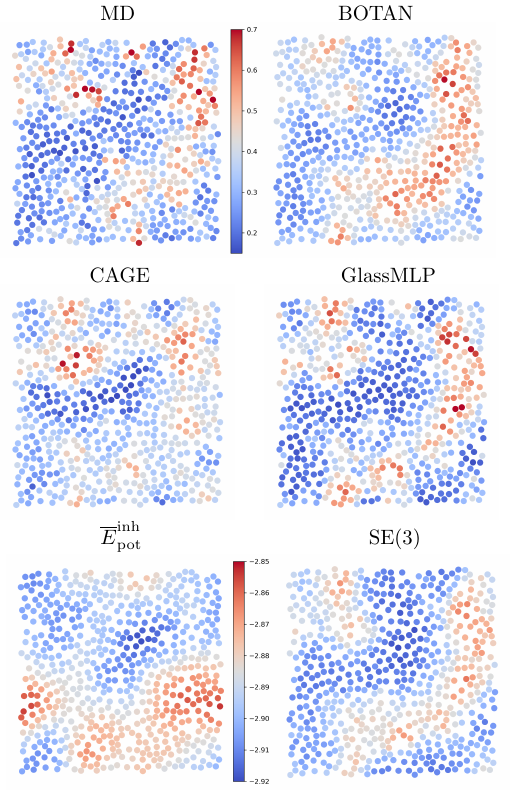}
    \caption{ Snapshots of an exemplary configuration of the KA2D system (shown are only type A particles). Each particle is coloured according to its propensity of displacement $\mathcal{R}$ at $t=\tau_\alpha$, as calculated from MD simulations or predicted from various ML techniques (see Fig.~3 in the main text for the respective Pearson correlations). Each snapshot has the same color bar, except $\overline{E}^\text{inh}_\text{pot}$. }
    \label{fig:KA_snap}
\end{figure}

The best way to visualise the predictions are snapshots of the amorphous structure, where each particle is coloured according to its propensity $\mathcal{R}.$ In Fig.~\ref{fig:KA_snap} we show an exemplary structure and the predicted propensities of displacement for various different models. Generally, strong correlations between the models and the MD simulations can be observed, as was expected from the high Pearson correlations. The strongest visible difference between MD and the ML models is that the latter are much smoother, in particular the ones based on GNNs (BOTAN and SE(3)).

\subsection{Cross correlations}

\begin{figure}
    \centering
    \includegraphics[scale=0.61]{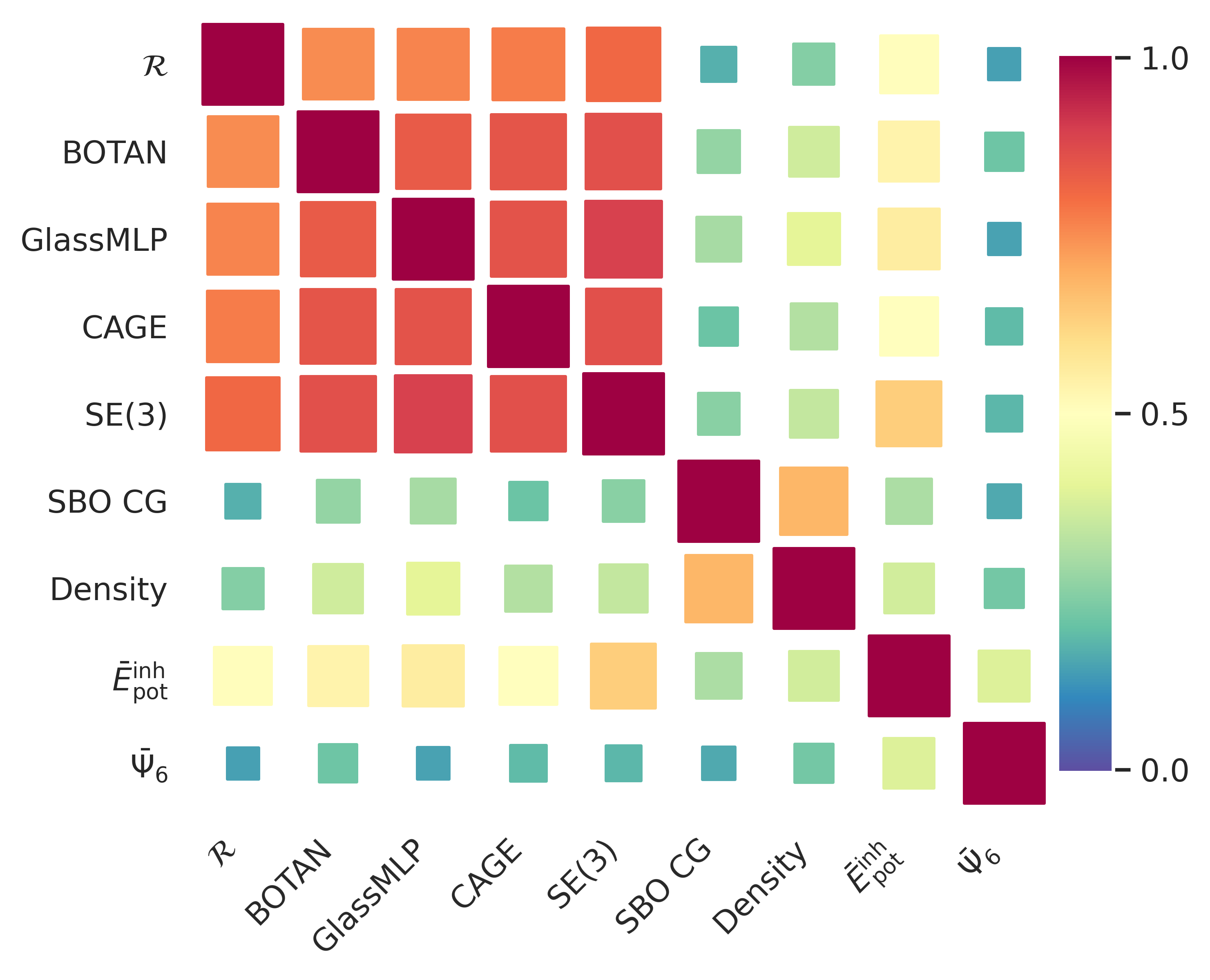}
    \caption{Pearson cross-correlations $\rho_\text{P}(X,Y)$ between different ML models and structural descriptors in the KA system (3D) at $t=\tau_\alpha$.}
    \label{fig:cross_correlations}
\end{figure}

To study connections between the different models, we also investigate the cross correlations between the different ML techniques and other structural descriptors. Unsurprisingly, the strongest cross correlations are between the various ML models, since they all correlate strongly with propensity. The strongest correlation can, in fact, be detected between SE(3) and GlassMLP (see Fig.~\ref{fig:cross_correlations}). This can be rationalized by the observation that both techniques also have the strongest correlation with the coarse-grained potential energy, $\overline{E}^\text{inh}_\text{pot}$. This could be expected since both techniques receive the potential energy as explicit input.

Another strong cross correlation is observable between the unsupervised SBO CG technique and density, showing that these two descriptors are strongly linked. This indicates that the strongest asymmetries in the structure emerge from density fluctuations.

\subsection{Learning curves}

\begin{figure}
    \centering
    \includegraphics[scale=0.9]{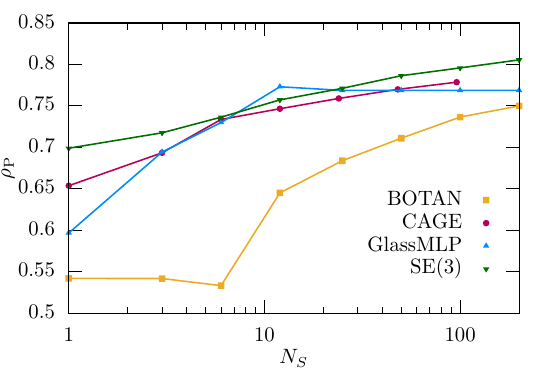}
    \caption{Performance of network, as quantified by the Pearson correlation coefficient $\rho_\text{P}$, trained with differently sized training sets in the KA system (3D) at $t=\tau_\alpha$. $N_S$ denotes the number of samples, where each sample consists of $N=3277$ particles.}
    \label{fig:learning_curve}
\end{figure}

An important feature of ML models is also how they cope with limited training data. In standard fitting procedures, one would usually expect that the more fitting parameters a model has, the more training data is required to achieve a certain accuracy. This 'rule of thumb' has, however, been repeatedly questioned for deep networks, which often perform well even when overparameterized. When investigating Fig.~\ref{fig:learning_curve} we observe that indeed SE(3) already performs excellent, even with only $N=3277$ data points, despite having $>50$k fitting parameters. Furthermore, the increase in performance is very continuous, implying that the network might perform even better with $>400$ independent structures (which was the maximum of the dataset available). The learning curves of CAGE and BOTAN closely resembles SE(3) just with a very small reduction in performance.

Contrarily, the learning curve of GlassMLP starts at a significantly smaller Pearson correlation, but quickly catches up with SE(3) to slightly surpass its performance at around $N_S=10$ (i.e. 32770 data points). However, the performance of GlassMLP does not further improve for larger $N_S > 20$. It might be necessary to increase the number of structural input descriptors to improve GlassMLP in situations where much training data is available. It should be mentioned that there is an important difference between the training procedures of SE(3) and the other models. While GlassMLP uses for $N_S=1$ only twice the number of epochs for training (i.e. $N_e=600$), since no improvement in performance can be observed beyond this point, SE(3) requires a significant increase in epochs for small $N_S$: $N_e(N_S) = 8 \cdot 10^4 / N_S.$ This implies that for $N_S=1,$ the training time of GlassMLP is roughly 5 seconds, while for SE(3) it remains roughly 11 hours.

\subsection{Transferability}

\begin{figure}
    \centering
    \includegraphics[scale=0.92]{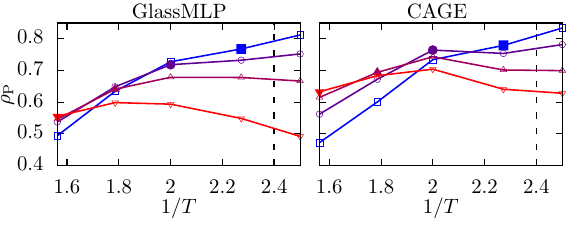}
    \caption{Transferability in temperature $T$ of trained networks in the KA system (3D) at the structural relaxation time $\tau_\alpha$.  The figures is identical to Fig.~6 in the main text, just featuring an additional temperature $T=0.4.$  }
    \label{fig:KA_transferability_appendix}
\end{figure}

We have investigated in detail the transferability of trained networks in the benchmarking section V of the main text. Here, we study for two models the performance of GlassMLP and CAGE when applied to an even lower temperature $T=0.4,$ taken from the dataset presented in Ref.~\cite{coslovich2018dynamic}. Different from the main KA dataset we only have $N_S=12$ individual structures available, but use $N_R=100$ different isoconfigurational replicas. Due to the larger number of replicas we expect the Pearson correlations to be slightly larger than for $N_R=32.$ The results in Fig.~\ref{fig:KA_transferability_appendix} indeed confirm this expectation, nevertheless, they also highlight the strong transferability of the models. A network trained at $T=0.44$, with a relaxation time an order of magnitude smaller than at $T=0.4$, CAGE can predict structural relaxation with a Pearson correlation of up to $\rho_\text{P} = 0.84$. We believe that this is an important result towards using transferability to study glassy liquids at extremely low temperature.

\subsection{Bond-breaking propensity in the KA system (3D)}

\begin{figure}[h]
    \centering
    \includegraphics[scale=0.95]{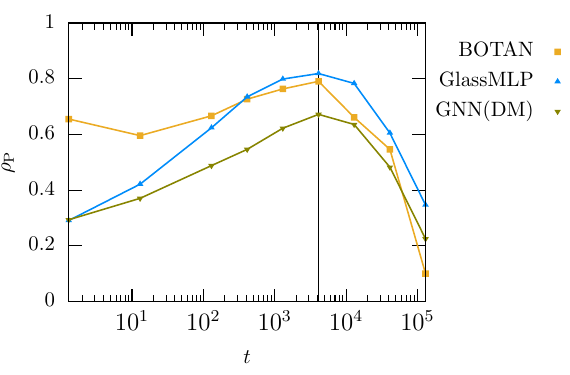}
    \caption{Pearson correlation $\rho_\text{P}$ with the ground truth in the KA system (3D) at $T=0.44$ for various different techniques. Different from Fig.~3, here the dynamical variable is the bond-breaking propensity. The vertical line marks the structural relaxation time $\tau_\alpha$. GNN(DM) refers to the original GNN \cite{bapst2020unveiling}. }
    \label{fig:KA_BB_pearson}
\end{figure}

Finally, we also briefly investigate the performance of ML models and other structural descriptors in predicting the bond-breaking propensity in the KA system (see Fig.~\ref{fig:KA_BB_pearson}). While the predictability for small times is much weaker than for the propensity of displacements, $\mathcal{R},$ analysed in Fig.~3, we observe better performance for longer times around the structural relaxation time scale $t = \tau_\alpha. $ This observation is identical to the discussion in Sec.~V of the main text after Fig.~5 for the bond-breaking propensity in the KA2D system. We further find that both BOTAN and GlassMLP significantly outperform the original GNN proposed by DeepMind (\cite{bapst2020unveiling}, GNN(DM)). This shows the great advancements in the field within the past three years. We hope that this roadmap further fuels the development and application of ML techniques to analyse glassy liquids.

%\FloatBarrier

\bibliography{library_local}